\documentclass[12pt]{article}
\pdfoutput=1
\usepackage{a4wide}
\usepackage[centertags]{amsmath}
\usepackage{amssymb}
\usepackage[sort&compress,numbers]{natbib}
\usepackage{ifpdf}
\usepackage{setspace}
\usepackage{amsfonts}
\usepackage{color}
\usepackage{threeparttable}
\usepackage{cancel}
\usepackage{tikz}
\usepackage{ulem}
\usetikzlibrary{decorations.pathmorphing}
\usepackage{graphicx}
\usepackage{subcaption}
\usepackage{verbatim}
\usepackage{xcolor}
\usepackage{bbm}
\usepackage{tikz}
\usepackage{pgfplots}
\pgfplotsset{compat=1.16}
\usepackage{tikz}
\usepackage{amsmath}
\usepackage{pgfplots}
\usepackage{amsmath}
\usepackage{amssymb}
\usepackage{amssymb}
\usepackage[sort&compress,numbers]{natbib}
\usepackage{ifpdf}
\usepackage{setspace}
\usepackage{amsfonts}
\usepackage{color}
\usepackage{threeparttable}
\usepackage{cancel}
\usepackage{tikz}
\usetikzlibrary{decorations.pathmorphing}
\usepackage{feynmf}
\usepackage[compat=1.1.0]{tikz-feynman}
\tikzfeynmanset{compat=1.1.0}
\usepackage{verbatim}
\usepackage{xcolor}
\usepackage{amsmath}
\usepackage{graphicx}
\usepackage{sidecap}
\usepackage{float}

\usepackage{yfonts}

\usetikzlibrary{decorations.markings}
\ifpdf
\usepackage[colorlinks=true,
linkcolor=black,
citecolor=black,
urlcolor=blue,
filecolor=blue,
pdfstartview=FitV,
pdftitle={},
pdfauthor={},
pdfsubject={Hydrodynamics},
pdfkeywords={hydrodynamics, AdS/CFT},
pdfpagemode=None,
bookmarksopen=true
]{hyperref}
\else
\usepackage[hypertex]{hyperref}
\fi
\usepackage{rotating}
\usepackage{rotating}
\makeatletter
\@addtoreset{equation}{section}

\makeatletter
\renewcommand\section{\@startsection {section}{1}{\z@}%
	{-3.5ex \@plus -1ex \@minus -.2ex}
	{2.3ex \@plus.2ex}%
	{\normalfont\large\bfseries}}
\renewcommand\subsection{\@startsection{subsection}{2}{\z@}%
	{-3.25ex\@plus -1ex \@minus -.2ex}%
	{1.5ex \@plus .2ex}%
	{\normalfont\bfseries}}

\def\App#1{Appendix \ref{#1}}

\title{Comparison between Causal and Acausal Diffusion:\\
	 a Schwinger-Keldysh Effective Field Theory Perspective}
     
\date{}

\author{Navid Abbasi$^{a,b,c}$\footnote{abbasi@lzu.edu.cn},  
\ Matthias Kaminski$^{d}$\footnote{mski@ua.edu},
\ Dirk H.\ Rischke$^{b,e}$\footnote{drischke@itp.uni-frankfurt.de}\\[2mm]
	\small{\textit{$^{a}$School of Nuclear Science and Technology, Lanzhou University,}}\\
	\small{\textit	{ 
			222 South Tianshui Road, Lanzhou 730000, China }} \\[2mm]
	\small{\textit{$^{b}$Institut f\"ur Theoretische Physik, Johann Wolfgang Goethe--Universit\"at,}}\\
	\small{\textit	{ 
			Max-von-Laue-Str.\ 1, D-60438 Frankfurt am Main, Germany}} \\[2mm]
	\small{\textit{$^{c}$ExtreMe Matter Institute EMMI,
			GSI Helmholtzzentrum f\"ur Schwerionenforschung,}}\\
	\small{\textit	{ 
			Planckstrasse 1,
			D-64291 Darmstadt, Germany}} \\[2mm]
    \small{\textit{$^{d}$  Department of Physics and Astronomy, University of Alabama, }}\\
	\small{\textit	{
    514 University Boulevard, Tuscaloosa, AL 35487, USA}} \\[2mm]
	\small{\textit{$^{e}$Helmholtz Research Academy Hesse for FAIR,}}\\
	\small{\textit	{ 
			Max-von-Laue-Str.\ 12, D-60438 Frankfurt am Main, Germany}} \\
}

\begin{document}
	
	\setlength{\baselineskip}{16pt}
	\begin{titlepage}
		\maketitle
		\vspace{-36pt}
		
		\begin{abstract}
In Fick’s time-honored theory of diffusion, the system responds instantaneously to external perturbations, resulting in acausal behavior. Maxwell-Cattaneo theory addresses this issue by introducing a relaxation time, rendering the diffusion process causal. We focus on systems where this relaxation time is comparable to the diffusion time and significantly larger than the relaxation times of all other non-conserved operators. 
In such systems, late-time diffusion is influenced by this relaxation process, leading to a theory of quasi-diffusion. Using the Schwinger-Keldysh Effective Field Theory (SK-EFT) framework, we compare the theories of diffusion and quasi-diffusion by analyzing the real-time dynamics of correlation functions both  in linear response and at one-loop order. In particular, we show that the one-loop corrections in the causal (quasi-diffusion) theory, in both the underdamped and the overdamped cases, are governed by two universal functions. In the overdamped case, the behavior mirrors that of the acausal (diffusion) theory, which has recently been applied for precision tests of SK-EFT in diffusive systems. We suggest that our results in the underdamped limit can also be used for precision tests in quasi-diffusive systems.

		\end{abstract}
		\thispagestyle{empty}
		\setcounter{page}{0}
	\end{titlepage}

	\renewcommand{\baselinestretch}{1}  
	\tableofcontents
	\renewcommand{\baselinestretch}{1.2}  
	\section{Introduction}
Thermalization via diffusive processes is a common aspect of macroscopic non-integrable systems. 
The phenomenological theory of diffusion, proposed by Fick \cite{Fick:1855}, successfully explains diffusion in numerous physical systems.  
This theory describes how the density of a conserved charge, initially distributed non-uniformly in the system, will eventually become uniform throughout the entire system. 

Despite its wide range of applications, there is a severe shortcoming of Fick's law: the instantaneous response of the density current with respect to a change in the density distribution. 
While this leads to serious inconsistencies in relativistic systems, even in non-relativistic systems it makes the very early-time predictions of the Fick's law unreliable.
The Maxwell-Cattaneo equation resolves this problem by introducing a relaxation time, $\tau$, for the response of the current \cite{Cattaneo:1958}. 
This transforms Fick's law of diffusion, which is a parabolic differential equation, to a hyperbolic differential equation. 
In this paper, we refer to these two equations as \textit{acausal} and \textit{causal} diffusion equations, respectively.\footnote{The equation of causal diffusion is in fact a M\"uller-Israel-Stewart (MIS)-type equation for diffusion \cite{Koide:2006ef}. 
 The MIS equation is only causal if the relaxation time is not too small. }

 The two diffusion theories mentioned above have been extensively discussed in the literature (see, for example, Refs.~\cite{Pu:2009fj, Gavassino:2024ufs, Bemfica:2020zjp, Jain:2023obu}).
 One point of view is that the causal theory can be understood as an ultraviolet (UV)-regulated version of the acausal theory. 
Under this perspective, both theories yield identical physical predictions in the infrared (IR) or hydrodynamic limit. Consequently, the relaxation time $\tau$, introduced within the causal framework, becomes bounded from below, and acts primarily as a UV regulator, having no essential impact on the observable physics in the IR regime.

However, in this work, we adopt a different approach. 
Suppose the diffusive system under investigation possesses a well-separated spectrum of modes. 
Naturally, the slowest among these modes corresponds to diffusion, accompanied by a hierarchy of faster, non-hydrodynamic modes. 
If we further assume that the slowest non-hydrodynamic mode has a decay rate comparable to that of the diffusion mode, accurately describing the diffusion process requires accounting for both the diffusion mode and this slowest non-hydrodynamic mode. 
This approach aligns closely with the concept of quasi-hydrodynamics, or Hydro+, as introduced in Refs.~\cite{Grozdanov:2018fic, Stephanov:2017ghc} (see also Ref.~\cite{Brattan:2024dfv}).
It is also supported by derivations of causal theories of dissipative hydrodynamics using kinetic theory as underlying microscopic theory~\cite{Denicol:2011fa,Denicol:2012cn}.
Within this framework, the relaxation time $\tau$ in the causal theory acquires direct physical significance, leading to an entanglement of IR physics with the dynamics associated with the slow non-hydrodynamic mode characterized by $\tau$. 

In this work, we aim to \textbf{compare} these two theories, here often referred to as Theory A (for the acausal version) and Theory B (for the causal version), within the framework of Schwinger-Keldysh Effective Field Theory (SK-EFT).
Motivated by the requirements of real experiments utilizing systems in both condensed-matter physics \cite{Brown:2019} and nuclear physics \cite{Du:2021zqz}, we focus on comparing the correlation functions of density fluctuations in the two theories. 
This comparison will proceed in two stages: first, within the linear-response regime, and then by examining the effects of nonlinearities on the physical observables.

\begin{table}[!htb]
	\label{table one}
\centering
\resizebox{\textwidth}{!}{%
		\begin{tabular}{|c|c|c|c|}
			\hline
			\hline
			&&  acausal & causal  \\
			&&A&B\\
			\hline
			\hline
			1 & constitutive &&\\
			&relation	&		$\textbf{J}+ D \,\boldsymbol{\nabla} n=\,0$ & $ \tau\partial_t \textbf{J}+\textbf{J}+ D \,\boldsymbol{\nabla} n=\,0$\\
			\hline
			2 & linear &&\\
			&equation & $\,\partial_t  n-\,D \nabla^2  n=\,0$  & $	\tau \partial_t^2  n +\,\partial_t  n-\,D \nabla^2  n=\,0$\\
			\hline
			3& linear &&\\
		&	mode(s)  & $\omega=\,-i D \textbf{k}^2$ & $\omega_{1,2}=\,-\frac{i}{2 \tau}\big(1\mp \sqrt{1 -4 \tau D \textbf{k}^2}\big)$ \\
			\hline
		4&	quadratic&&\\
			& Lagrangian	&  $\mathcal{L}_2=iT \sigma(\nabla n_a)^2- n_a(\partial_t n-D \nabla^2 n)$\ &  $\mathcal{L}_2=iT \sigma(\nabla n_a)^2- n_a(\tau \partial_t^2n+\partial_t n-D \nabla^2 n)$\\
			 \hline
			 	5&	spectral &&\\
			 & function	&  $G^{(0)}_{{nn}_A}(\omega, \textbf{k})=\displaystyle\frac{2 T \chi D \textbf{k}^2}{\omega^2+(D \textbf{k}^2)^2}$ &  $G^{(0)}_{{nn}_B}(\omega, \textbf{k})=\displaystyle\frac{2 T \chi D \textbf{k}^2}{\omega^2+(\tau \omega^2-D \textbf{k}^2)^2}$\\
			 \hline			 
			 6&	nonlinear &&\\
			 & equation			& $\,\partial_t  n-\, \nabla^2 \big(D\, n+\frac{\lambda_D}{2} n^2+\frac{\lambda'_D}{6} n^3\big)=0$ & $\tau \partial_t^2 n +\partial_t  n- \nabla^2 \big(D n+\frac{\lambda_D}{2} n^2+\frac{\lambda'_D}{6} n^3\big)=0$\\
			 \hline
			 
		7&	 cubic &&\\ 
		&	Lagrangian& $\mathcal{L}_3=\,iT\chi \lambda_{\sigma}\,n(\nabla n_a)^2+\frac{\lambda_D}{2}\, \nabla^2	n_a\, n^2$ &$\mathcal{L}_3=\,\frac{\lambda_D}{2}\, \nabla^2	n_a\, n^2$ \\
			\hline
		8&	response &&\\
		&	function & $	G_{R_A}(\omega,\textbf{k})=\,\displaystyle \frac{i\,\sigma\,\textbf{k}^2}{\omega +i [D+\delta D(\omega,\textbf{k})] \textbf{k}^2}$ & $	G_{R_B}(\omega,\textbf{k})=\,\displaystyle\frac{i\,\sigma\,\textbf{k}^2}{- i [\tau+\delta \tau(\omega,\textbf{k})] \omega^2+ \omega +i [D+\delta D(\omega,\textbf{k})] \textbf{k}^2}$\\
			\hline
					9&one-loop & $	\delta D(\omega,\textbf{k})=\displaystyle\frac{\lambda_D^2 T \chi}{4D^{3/2}}(- i \omega)  F_d(\omega, \textbf{k})$  & $	\delta D(\omega,\textbf{k})=\displaystyle\frac{\lambda_D^2 T \chi}{16D^{3/2}}(- i \omega)F_d(\omega, \textbf{k}) $  \\
 		& correction &$	$& $\,\,\,\,\,\,\,\,\,\,\,\,\,\,\,\,\,\,\,\,\,\,\,\,\,\,	\times\left[2+\tau D \textbf{k}^2-\tau \omega(3i+\tau \omega)\right]^2$\\
 		               &&&$\delta \tau(\omega, \textbf{k})=0\,\,\,\,\,\,\,\,\,\,\,\,\,\,\,\,\,\,\,\,\,\,\,\,\,\,\,\,\,\,\,\,\,\,\,\,\,\,\,\,\,\,\,\,\,\,$\\
 		\hline 
 		10& non-analytic &  $F_1(\omega, k)=\big[D k^2-2i\omega\big]^{-1/2} $  & $F_1(\omega, k)=\big[D k^2-i\omega  (2-i\tau  \omega)\big]^{-1/2}$\\
 		& part &&\,\,\,\,\,\,\,\,\,\,\,\, \,\,\,\,\,\,\,\,\,\,\,$\times\big[ \tau D k^2+(1 -i \tau \omega)^2\big]^{-3/2}$\\
 		\hline
 		11 & pinch & $\tilde{\omega}= \displaystyle-\frac{i}{2}D \textbf{k}^2$ & $\tilde{\omega}_{11,22}=-\displaystyle\frac{i}{\tau}\big(1\mp\sqrt{1-\tau D \textbf{k}^2}\big)$ \\
 		 		& singularity &  & $\tilde{\omega}_{12,21}=-\displaystyle\frac{i}{\tau}\pm \frac{\sqrt{\tau D \textbf{k}^2 }}{\tau}\,\,\,\,\,\,\,\,\,\,\,\,\,\,\,\,\,\,\,\,\,$ \\
			\hline
		\end{tabular}%
	}
		\caption{Comparison between Theory A (acausal diffusion) and Theory B (causal diffusion). Results associated with Theory A have been found in Ref.~\cite{Chen-Lin:2018kfl} (Eq.~(9-A) is not explicitly given in this reference). Results associated with Theory B have been found in Ref.~\cite{Abbasi:2022aao}. Equations (4-B) and (5-B) have been reproduced in Ref.~\cite{Jain:2023obu}.)}
        	\label{Table}
\end{table}

The correlation functions can be calculated in the framework of SK-EFT. 
The (Fourier-space) results for the density-density correlation functions associated with the above two theories have been recently found in Refs.~\cite{Chen-Lin:2018kfl,Abbasi:2022aao}. 
Table \ref{Table} reviews these results. 
In the following, we refer to the pieces of information in the table by the simple notation ``Equation~(m-A)'' or ``Equation~(m-B)'', respectively. 
Here ``m" denotes the row number in the table, while A and B corresponds to the result for the acausal and causal theory given in the third and fourth columns, respectively. 
Note that throughout this paper, while our linear-response results are valid in arbitrary dimensions, we focus on the nonlinear (one-loop) results in $d=1$ dimension, as presented in Eqs.~(10-A) and (10-B). 
However, nonlinear results such as Eqs.~(11-A) and (11-B), which do not arise from loop-momentum integrations, remain valid in any number of spatial dimensions.

The goal of this work is to introduce quantitative measures for analyzing experimental data associated with the above mentioned two types of diffusive systems; those with only a diffusive mode at late time against the diffusive systems with an extra slow mode, with a frequency comparable with that of the diffusive mode.
To achieve this, we utilize the results of the EFT framework to calculate several physical observables. 
These include the real-time correlation function, the spectrum of excitations, the corrected diffusion constant, and the dynamic susceptibility.
By comparing the behavior of these quantities in the two theories, we aim to identify which of the two theories is more applicable to a given set of experimental data in a diffusive system.

Note that the EFT results are to be finally applied to specific physical systems. 
We assume that the latter have a length scale, $\ell$, and an energy scale, $\Delta$. 
Then all physical quantities in our EFT can be made dimensionless by using suitable combinations of $\ell$ and $\Delta$. 
For instance, in Ref.~\cite{Brown:2019}, $\ell$ corresponds to the lattice spacing, and $\Delta$ shows the strength of coupling in the Hubbard model. 
Using the same idea, we express all quantities by dimensionless numbers. 
Then for instance, $t=1$ corresponds to $t=1 \times \frac{\hbar}{\Delta}$; $x=0.5$ corresponds to $x=0.5 \ell$,  etc. In all our plots, we take the bare value of $D=1$, which is in fact $D=\frac{\ell^2\Delta}{\hbar}$. 
Let us note that we present our results in the rest frame of the system. 
\section{Constitutive relation and linear dynamics}
\begin{figure}
	\centering
	\begin{tikzpicture}
		\draw[line width=1.2pt,->, -{Stealth}] (-4,0) -- (4.5,0);
		\draw[draw,line width=1.2pt] (-3.5,-.1) -- (-3.5,0.1);
		\draw[draw,line width=1.2pt] (-0.5,-.1) -- (-0.5,0.1);
		\draw[draw,line width=1.2pt] (2.5,-.1) -- (2.5,0.1);
		\node[] at (-3.5,-.3) {$0$};
						\node[] at (-2.5,-.3) {$k^*$};
				\node[] at (-0.5,-.3) {$k_c$};
		\node[] at (4.7,0) {$k$};
				\node[] at (2.5,-.3) {$k_c^{\prime}$};

			\draw[fill=blue!30, draw=none, opacity=0.7] (-3.5,-1.5) rectangle (-2.5,-1.0);
	\draw[fill=magenta!30, draw=none, opacity=0.7] (-3.5,-1.0) rectangle (-2.5,-0.5);
		\draw[fill=gray, draw=none, opacity=0.4] (-3.5,-2.0) rectangle (-0.5,-1.5);
		\draw[fill=blue!30, draw=none, opacity=0.5] (-2.5,-1.5) rectangle (-0.5,-1.0);
		\draw[fill=magenta!30, draw=none, opacity=0.5] (-2.5,-1.0) rectangle (-0.5,-0.5);
		\node[magenta] at (-3.0,-0.8) {\textbf{A}};
		
		\draw[fill=gray, draw=none, opacity=0.4] (-0.5,-2.0) rectangle (2.5,-1.5);
		\draw[fill=blue!30, draw=none, opacity=0.5] (-0.5,-1.5) rectangle (2.5,-1.0);
		\node[blue] at (-1,-1.3) {\textbf{B}};
		
		\draw[fill=gray, draw=none, opacity=0.4] (2.5,-2.0) rectangle (4,-1.5);
		
		\draw[draw,dashed,line width=0.9pt] (-3.5,1.2) -- (-3.5,.2);		
				\draw[draw,dashed,line width=0.9pt] (-2.5,1.2) -- (-2.5,.2);		
					\node[magenta] at (-3.,.8) {$\omega$};
				\node[blue] at (-1.2,.8) {$\omega_1$};
		\draw[draw,dashed,line width=0.9pt] (-0.5,1.2) -- (-0.5,.2);
		\draw[draw,dashed,line width=0.9pt] (2.5,1.2) -- (2.5,.2);
			\draw[blue,draw,line width=.5pt,->,-{Stealth}] (-.9,.8) -- (-.5,.8);
		\node[blue] at (-1.2,.8) {$\omega_1$};
		\draw[blue,draw,line width=.5pt,->,-{Stealth}] (-.1,.8) -- (-.5,.8);
		\node[blue] at (.2,.8) {$\omega_2$};
					\draw[blue,draw,line width=.5pt,->,-{Stealth}] (2.1,.8) -- (2.5,.8);
\node[blue] at (1.8,.8) {$\omega_2$};
\draw[gray,draw,line width=.5pt,->,-{Stealth}] (2.9,.8) -- (2.5,.8);
\node[gray] at (3.2,.8) {$\omega_3$};
	\end{tikzpicture}
	\caption{
		Theory A is designed to work at small $k$, $k\lesssim k^*$, focusing on diffusive processes with a single mode $\omega$. At higher momenta, Theory A requires modifications through higher derivatives.
		This all may be useful up to a cutoff, beyond which diffusion is not the dominant process. To specify the cutoff and model the ``new physics" beyond it, a simple approach is to assume that the microscopic system exhibits a hierarchy of fast modes: $|\omega_1|\lesssim|\omega_2|\ll |\omega_3|\ll \ldots$ where $\omega_j(k\rightarrow0)\ne0$ ($j=2,3,\ldots$).
		 The modified version of Theory A, referred to as Theory B, introduces two modes; a diffusive mode $\omega_1$ which coincides with $\omega$ of Theory A at small momenta, and a fast mode $\omega_2$, the slowest among the fast modes. Now $\omega_1$ is a resummation of $\omega$ at high momenta, and the cutoff of Theory A corresponds to the momentum $k_c$ at which $\omega_2$ collides with $\omega_1$, and $D k_c^2=1/2\tau$.}
	\label{New_physics}
\end{figure}
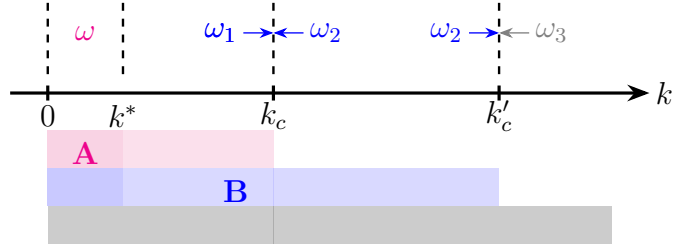

Acausal diffusion is simply the conservation of a density $n$, whose flux is phenomenologically described by a current $J$, expressed in terms of the spatial gradient(s) of the density, see Eq.~(1-A) in Table~\ref{Table}.
The dynamics is then simply governed by Eq.~(2-A), and describes the late-time diffusion of the conserved density $n$ over large distances, where gradients are small.
In Fourier space, it features a single characteristic mode, given by Eq.~(3-A), which is valid in the small-momentum limit, i.e. $k\lesssim k^*$ (see Fig.~\ref{New_physics}).

In practice, the momentum $k$ can exceed $k^*$. 
Although this extension is mathematically straightforward, it leads to physical inconsistencies. 
The group velocity associated with the diffusive mode, namely $\big|\frac{\mathrm{d}\omega}{\mathrm{d} k}\big|\sim D k$, can grow without bound at large momenta, potentially violating causality. 
This issue is not merely a mathematical artifact; it indicates that some form of ``new physics" must emerge at higher momenta, for example beyond a characteristic scale $k_c$, which acts as the effective cutoff of Theory A and restores causal behavior. 

The new physics beyond the cutoff cannot be fully described by Eqs.~(1-A) and (2-A) in Table~\ref{Table}. In a class of diffusive systems with a well-separated spectrum of non-hydrodynamic modes, the physics beyond the cut-off is described by Theory B. 
As illustrated in Fig.~\ref{New_physics}, Theory B, which is characterized by $\omega_1$ and $\omega_2$, serves as an extrapolation of Theory A.
This way, the slowest non-hydrodynamic mode is $\omega_2$ and the diffusion mode is $\omega_1$, which is the corrected version of the diffusion mode in Theory A via incorporating higher-order terms ($k^4, k^6, \ldots$).

\begin{figure} [tb]
	\centering
	\centering
	\includegraphics[width=0.45\textwidth]{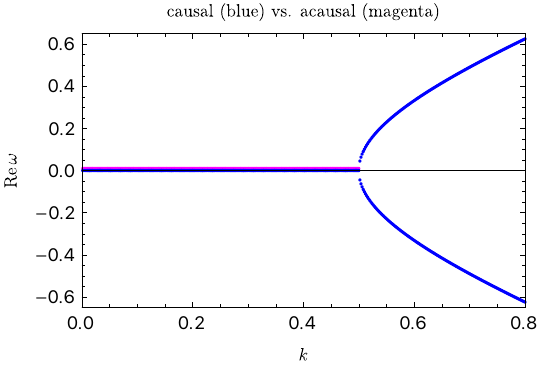} \,\,\,	\includegraphics[width=0.45\textwidth]{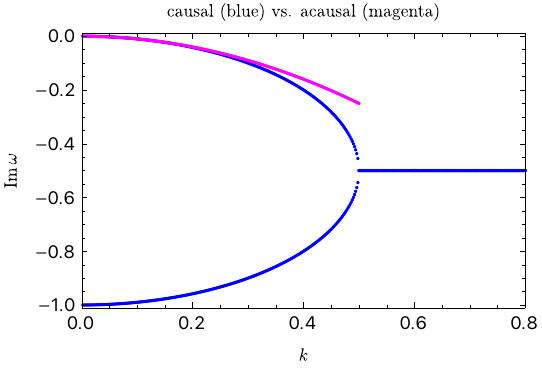} \\
	\caption{The dispersion relation of Theory A can, at best, be ``mathematically" applied up to the scale where new physics emerges. This corresponds to the collision of the diffusive mode with the fast mode in Theory B, which occurs at $k=0.5$ in this case. (We have set $D=1$ and $\tau=1$.) However, the ``physical" validity of Theory A is not ensured for the entire range $k<0.5$. As discussed in the text, Theory A remains ``physically" valid only in the small-momentum regime, where the magenta curve coincides with the blue curve. The range of validity of Theory A is $k<k^*\sim 0.25$ in this case. }  
	\label{spectrum_linear}
\end{figure}

Theory B can be taken as Maxwell-Cattaneo theory (see Ref.~\cite{Romatschke:2017ejr} for a comprehensive discussion). 
The decisive feature is the introduction of a relaxation time $\tau$ in the current equation (see Eq.~(1-B) in Table~\ref{Table}). 
This leads to the following key distinctions:
\begin{itemize}
	\item The two theories yield identical predictions only at small momenta, $k\lesssim k^*$.
	\item The naive cutoff of Theory A can be identified as the momentum at which $\omega_1$ and $\omega_2$ from Theory B collide (see the collision point in the right panel of Fig.~\ref{spectrum_linear}).
	\item Below the naive cutoff, $\omega_1$ represents the resummation of $\omega$, allowing it to remain the diffusive mode even beyond the range of validity of Theory A, i.e., for  $k^*\lesssim k  <k_c$.\footnote{See Ref.~\cite{Heller:2020uuy} for a discussion on $k_c$ in Theory B, or the so-called Telegrapher’s equation, in the linear response regime. 
See Ref.~\cite{Brattan:2024dfv} for a modified Maxwell-Cattaneo equation, which describes systems in motion with a constant velocity. 
See also Refs.~\cite{Dash:2022xkz,Lier:2025wfw},
 where the Telegrapher's equation is used to perform numerical simulations in an MHD setup.} 
 
    \item Unlike Theory A, Theory B can describe causal diffusion, as Eq.~(3-B) in Table~\ref{Table} is a hyperbolic equation. 
    Causality is ensured by demanding $D\le \tau$ \cite{Pu:2009fj,Grozdanov:2018fic,Romatschke:2017ejr,Romatschke:2009im}.
    \item Theory B can be used above $k_c$, up to $k_c^{\prime}$. 
    The location of $k_c^{\prime}$ depends on the scale where other new physics emerges beyond $k_c$, that we do not discuss here. 
    For a more detailed discussion, see Ref.~\cite{Denicol:2011fa}.
	\end{itemize}
The difference in the spectrum of linear excitations is illustrated in Fig.~\ref{spectrum_linear}. The corresponding modes are given by Eqs.~(3-A) and (3-B) in Table \ref{Table}. 
Setting $\tau=1$ and $D=1$, the cutoff of Theory A is $k_c=0.5$. For this reason, we present the real and imaginary parts of the dispersion relation in Theory A only within the range $k<0.5$.

As seen in the right panel of Fig.~\ref{spectrum_linear}, Theory B, by design, possesses two characteristic modes, enabling it to access a broad range of frequencies, whereas Theory A operates only at low frequencies. This distinction has significant implications for real-time dynamics. Theory B remains valid on timescales of order $\tau$, while Theory A is applicable only at late times, when $\tau \ll t$.

In other words, Theory A cannot consistently account for early-time dynamics, particularly when the system's thermalization time is not short. In this situation, Theory B is the appropriate framework for describing the dynamics.
\section{Effective field theory without interactions}
The two equations (2-A) and (2-B) in Table~\ref{Table} capture the dissipative effects in the system. 
In order to incorporate the effect of fluctuations, one systematic way is to construct the SK-EFT for density fluctuations.\footnote{See Refs.~\cite{Liu:2018kfw,Glorioso:2017fpd,Jensen:2018hhx,Jensen:2018hse,Haehl:2018lcu,Grozdanov:2013dba, Mullins:2025vqa} for constructing the SK-EFT. 
See also Ref.~\cite{Basar:2024srd} and references therein for recent developments of fluctuating hydrodynamics and SK-EFT.}
\subsection{A quick review of known results}
 The associated quadratic Lagrangian, which corresponds to the linear dynamics, is given by Eqs.~(4-A) \cite{Chen-Lin:2018kfl} and (4-B) \cite{Abbasi:2022aao} in Table~\ref{Table} for the two theories. 
 Note that $\sigma= \chi D$ is the conductivity, and $\chi$ is the susceptibility. 
 It is clear that the classical equation of motion of the density field $n$ derived from Eqs.~(4-A) and (4-B) in Table~\ref{Table} is given by Eqs.~(2-A) and (2-B), respectively. 
 The field $n_{a}$ is the noise field.

  Equation~(4-B) has been reproduced in Ref.~\cite{Jain:2023obu} (see also Ref.~\cite{Mullins:2023ott}). 
 The analogue of Eq.~(4-A) in systems with a gravity dual was found in Ref.~\cite{Glorioso:2018mmw} (see also Ref.~\cite{Bu:2021clf}). 
 Recently, Ref.~\cite{Liu:2024tqe} constructed Eq.~(4-B) in a holographic setup (see also Ref.~\cite{Ahn:2025odk}).

	%

\subsection{Real-time correlation function: $G^{(0)}_{nn}(t,\mathbf{k})$}
\begin{figure} 
	\centering
	\centering
	\includegraphics[width=0.44\textwidth]{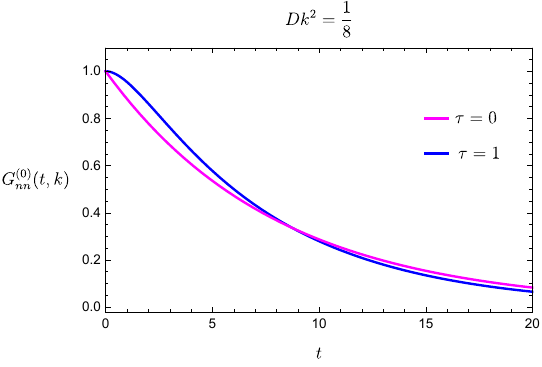}\,\,\,\,\includegraphics[width=0.44\textwidth]{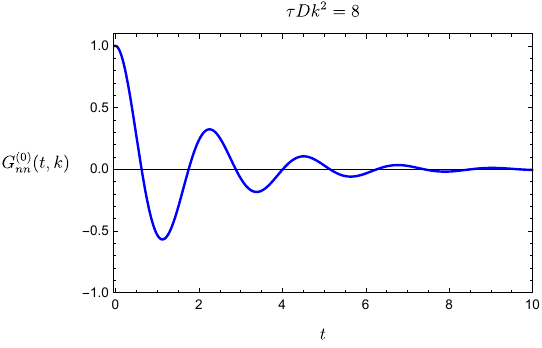} \\
	\caption{At any $k$,  $G_{{nn}_{A}}^{(0)}(t,\mathbf{k})$ decays exponentially (magenta curve, $\tau =0$)). 
    However, depending on the value of $\tau D k^2$, $G_{{nn}_{B}}^{(0)}(t,\mathbf{k})$ shows either an \textit{underdamped} ($\tau D k^2>1/4$) or an \textit{overdamped} ($\tau D k^2<1/4$) behavior (blue curves). }
      
	\label{correlation_linear}
\end{figure}

Using standard techniques of field theory, one can derive the correlation functions of the density field, i.e., $G^{(0)}_{nn}(\omega,\mathbf{k})$, from $\mathcal{L}_2$.  
Here, the superscript $(0)$ stands for  the free EFT (without interaction). 
In real experiments, one commonly measures the real-time correlation functions for a fixed wave number. 
The latter could correspond to the wave number of external sources that perturb the system \cite{Brown:2019}. 
For this reason, in this subsection we compare Theories A and B by considering   $G_{{nn}_{A/B}}^{(0)}(t,\mathbf{k})=\int\frac{\mathrm{d}\omega}{2\pi}G_{{nn}_{A/B}}^{(0)}(\omega,\mathbf{k})e^{- i \omega t}$,
with the integrand given by Eqs.~(5-A) and (5-B), respectively. 
For $t>0$, by picking up the residue of $G_{{nn}_{A/B}}^{(0)}(\omega,\mathbf{k})$ at the pole(s) lying in the lower half of the complex $\omega$-plane, we find:
\begin{align}\label{G_nn_A}
		G_{{nn}_{A}}^{(0)}(t,\mathbf{k})&=\,T \chi\,e^{-D k^2 t}\;,\\\label{G_nn_B}
		G_{{nn}_{B}}^{(0)}(t,\mathbf{k})&=\,T \chi\,\frac{2\tau D k^2}{\sqrt{1- 4 \tau D k^2}}\left[\frac{e^{\sqrt{1- 4 \tau D k^2}\frac{t}{2\tau}}}{1-\sqrt{1- 4 \tau D k^2}}-\frac{e^{-\sqrt{1- 4 \tau D k^2}\frac{t}{2\tau}}}{1+\sqrt{1- 4 \tau D k^2}}\right]e^{-\frac{t}{2\tau}}\;.
\end{align}  
 Figure~\ref{correlation_linear} compares this function for Theories A and B.
\\
In Theory A (where $\tau=0$), consistent with Eq.~\eqref{G_nn_A}, $G_{nn}^{(0)}(t,\mathbf{k})$ always decays exponentially with time. 
For Theory B, depending on the value of $\tau D k^2$, $G_{{nn}_{B}}^{(0)}(t,\mathbf{k})$ shows two qualitatively different behaviors:
\begin{enumerate}
			\item For $1-4 \tau D k^2>0$ the contribution associated with the two terms in the brackets in Eq.~\eqref{G_nn_B} is decaying; the correlation is therefore ``\textit{overdamped}", 	
			\begin{align}\label{over_damped}
			\text{overdamped}: \,\,\,\,\,\,\,\,	G_{{nn}_{B}}^{(0)}(t,\mathbf{k})=\,T \chi\left[\frac{1}{\alpha}\sinh\left(\alpha \frac{t}{2\tau}\right)+\cosh\left(\alpha \frac{t}{2\tau}\right)\right]e^{-\frac{t}{2\tau}}\;,
			\end{align}  
				with $\alpha = \sqrt{1 - 4 \tau Dk^2}$. Obviously, when $\tau\rightarrow 0$, the above expression reduces to Eq.~\eqref{G_nn_A}.
			This corresponds to the range of small $k$ values  in Fig.~\ref{spectrum_linear}, where the two modes are purely imaginary.
	\item For $1-4 \tau D k^2<0$ the contribution associated with the two terms in the brackets in Eq.~\eqref{G_nn_B} are oscillatory and both are damped with the factor $e^{-t/2\tau}$. 
    The correlation is therefore ``\textit{underdamped}",
	\begin{align}\label{under_damped}
				\text{underdamped}: \,\,\,\,\,\,\,\,	G_{{nn}_{B}}^{(0)}(t,\mathbf{k})=\,T \chi\left[\frac{1}{\beta}\sin\left(\beta \frac{t}{2\tau}\right)+\cos\left(\beta \frac{t}{2\tau}\right)\right]e^{-\frac{t}{2\tau}}\,\,\,\,,
	\end{align}  
	with $\beta=\sqrt{4 \tau D k^2-1}$. 
	This corresponds to the range of large $k$ values in Fig.~\ref{spectrum_linear}, where the two modes have a non-vanishing real part.
	\end{enumerate}
Note that both of the above two behaviors have been found in Ref.~\cite{Kapusta:2014dja}, and also have been observed in experiment, see, e.g., Fig.~2 in Ref.~\cite{Brown:2019}.\footnote{Reference~\cite{Brown:2019} measures $n(t,k)$ in an ultra-cold gas of $^6\text{Li}$ on a two-dimensional lattice. 
Figure~2 in this reference reports the result of this measurement for some specific values of $k$. 
On the other hand, we know that $G_{nn}(t,\mathbf{k})$ follows exactly the same equation that $n(t,k)$ satisfies (see Appendix \ref{App_Landau}). 
Then one naturally expects  Eqs.~\eqref{over_damped} and \eqref{under_damped} to be consistent with the data associated with the $n(t,\mathbf{k})$ measurement in Ref.~\cite{Brown:2019}.}

Another manifestation of causality which distinguishes Theory B from Theory A is obviously observed in the behavior of $G_{{nn}_{A/B}}^{(0)}(t,\mathbf{k})$ in the vicinity of $t=0$. 
Extending Eqs.~\eqref{G_nn_A} and \eqref{G_nn_B} to $t<0$,\footnote{Note that $G_{nn}(t,\mathbf{k})=G_{nn}(-t,-\mathbf{k})$, and $G_{n J_i}(t, \mathbf{k})=G_{J_i n}(-t, -\mathbf{k})$. 
See Appendix~\ref{App_Landau} for details.} one finds the following property:
\begin{eqnarray}\label{G_nn_A_partial}
	\partial_t G_{{nn}_{A}}^{(0)}(t=0^{\pm},\mathbf{k})&=&\,\mp Dk^2T \chi\;,\\\label{G_nn_B_partial}
	\partial_t G_{{nn}_{B}}^{(0)}(t=0,\mathbf{k})&=&\,0\;.
\end{eqnarray}  
\begin{SCfigure}
	\centering
	\includegraphics[width=0.45\textwidth]{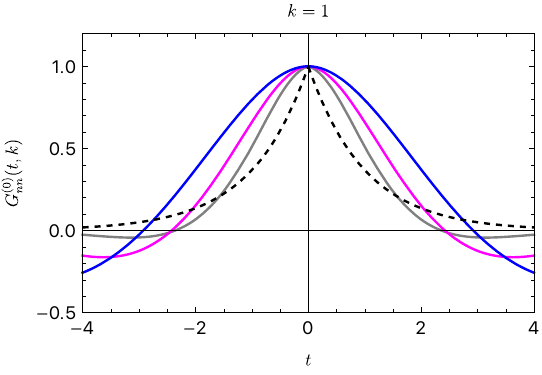}	
	\caption{The derivative of the dashed curve (corresponding to the acausal theory with $D k^2=1$) is  discontinuous at $t=0$, while the inertial effect prevents such a discontinuity in the causal case. The curves in gray, magenta, and blue correspond to $\tau D k^2=0.5, 1, 2$, respectively.\\
		\\
	}  
	\label{singularity}
\end{SCfigure}
This is shown in Fig.~\ref{singularity}. 
The fact that the derivative is not continuous is the sign of an acausality. 
This can be understood in a way independent of the SK-EFT results. 
Taking into account the conservation equation $\partial_t n +\nabla \cdot J=0$, the correlation function $G_{{nn}}^{(0)}(t,\mathbf{k})$ satisfies the following equation (see Appendix \ref{App_Landau})
\begin{equation}\label{conservation_G_nn}
\partial_t G^{(0)}_{nn}(t,\mathbf{k})+ ik_j G^{(0)}_{J_jn}(t,\mathbf{k})=0\;,\,\,\,\,t>0\,.
\end{equation}  
\begin{itemize}
	\item In Theory A, Eq.~(1-A) leads to $G^{(0)}_{J_jn}(t,\mathbf{k})+ i k_j G^{(0)}_{nn}(t,\mathbf{k})=0$. 
   Initializing at $t=0$, in order to integrate Eq.~\eqref{conservation_G_nn} one needs only a single boundary condition, that is $G^{(0)}_{nn}(t=0,\mathbf{k})=T \chi$ \cite{Landau_1}. 
    Since $G_{nn}(t,\mathbf{k})$ is an even function by definition, it is required that $\partial_tG^{(0)}_{nn}(t=0^{+},\mathbf{k})+\partial_tG^{(0)}_{nn}(t=0^{-},\mathbf{k})=0$. 
    But this does not determine the value of right and left derivatives at $t=0$. 
    
    Following the above discussion, Eq.~\eqref{conservation_G_nn} can be recast in the form $\pm\partial_t G^{(0)}_{nn}(t^{\pm},\mathbf{k}) + D k^2  G^{(0)}_{nn}(t^{\pm},\mathbf{k})=0$, for the two sides of $t=0$. The right and left derivatives given in Eq.~\eqref{G_nn_A_partial} then follow directly from imposing the boundary condition  $G^{(0)}_{nn}(t=0,\mathbf{k})=T \chi$ on this equation. In contrast, in Theory B, as we will see below, the left and right derivatives at $t=0$ must vanish. This follows from the fact that the cross-correlator between the current and the density is required to vanish at $t=0$, thereby forbidding any instantaneous response of the current to density gradients. In Theory A, however, this correlator can be nonzero at $t=0^{\pm}$.
	\item In Theory B, $G^{(0)}_{J_jn}(t,\mathbf{k})$ follows a first-order differential equation, constructed from Eq.~(1-B):
	\begin{equation}\label{equation_G_jn}
		\tau\partial_t G^{(0)}_{J_jn}(t,\mathbf{k})+ G^{(0)}_{J_jn}(t,\mathbf{k})+ i D k_j G^{(0)}_{nn}(t,\mathbf{k})=0\;.
	\end{equation}  
	Solving Eqs.~\eqref{conservation_G_nn} and \eqref{equation_G_jn} requires two boundary conditions:
	\begin{eqnarray}\label{boundary_condition}
G^{(0)}_{nn}(t=0,\mathbf{k})&=&T \chi\;,\\\label{boundary_condition_2}
G^{(0)}_{J_jn}(t=0,\mathbf{k})&=&0\;.
	\end{eqnarray}  
	While the first one is expected on account of thermal fluctuations, the second is fixed by requiring that the  demanding that the cross-correlator $G^{(0)}_{J_i n}(t=0, \textbf{k})$ be well-defined. 
    Then, the symmetry property of this correlator, namely $G_{J_i n}(t, \textbf{k})=-G_{J_i n}(-t, \textbf{k})$ \cite{Landau_1}, implies $G^{(0)}_{J_i n}(t=0, \textbf{k})=0$.
    Substituting the latter into Eq.~\eqref{conservation_G_nn}, we recover Eq.~\eqref{G_nn_B_partial}.
	\end{itemize}
\section{Nonlinear effects: Effective field theory with interactions}
The coefficients $\sigma$, $D$, and $\tau$ in the quadratic Lagrangian (see Eqs.~(4-A) and (4-B) in Table~\ref{Table}) could be in general functions of density fluctuations. 
At the level of the equations of motion, this leads to nonlinearities and the equation of motion will no longer be linear.
One can expand these coefficients around the background value of the density, $n=\bar{n}+\delta n$,
\begin{equation}\label{coupling}
D(n)=D+D'\delta n+\frac{1}{2}D''\delta n^2+\ldots  \;,
\end{equation}  
where $D$, $D'$, $D''$, etc. are evaluated at $\bar{n}$.
In Table~\ref{Table}, $\lambda_D\equiv D'$, and $\lambda_D'\equiv D''$.
 These are Wilsonian coefficients of EFT. 
 The coefficient $D$ appears just in $\mathcal{L}_2$ and consequently in the linear response.  
 The coefficients $D'$, $D''$, etc. give rise to interaction terms in the Lagrangian, contributing to $\mathcal{L}_3$ and higher orders (see Eqs.~(7-A) and (7-B) in Table~\ref{Table}). 
   In the following subsections, we study several predictions of EFT associated with the one-loop correction to the correlation function (see Appendix~\ref{loop}). 
   As we will discuss, in the overdamped regime this yields the dominant nonlinear contribution across all diffusive systems. 
   In contrast, in the underdamped limit, the one-loop contribution represents the leading nonlinear effect only in a specific subset of diffusive systems—particularly those characterized by a large susceptibility.

\subsection{Spectrum}
\begin{figure}[H]
	\centering
	\centering
	\includegraphics[width=0.4\textwidth]{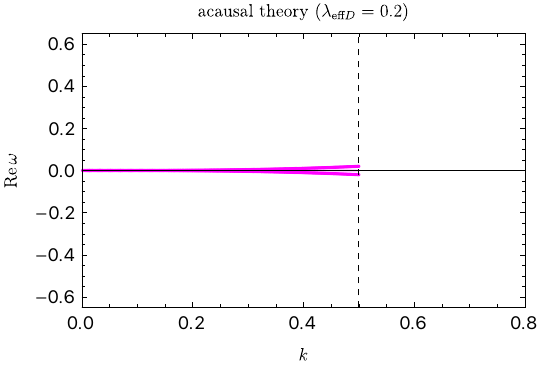}\,\,\includegraphics[width=0.4\textwidth]{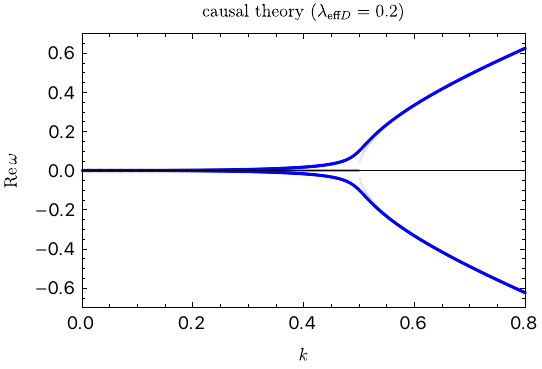}\\
    \includegraphics[width=0.4\textwidth]{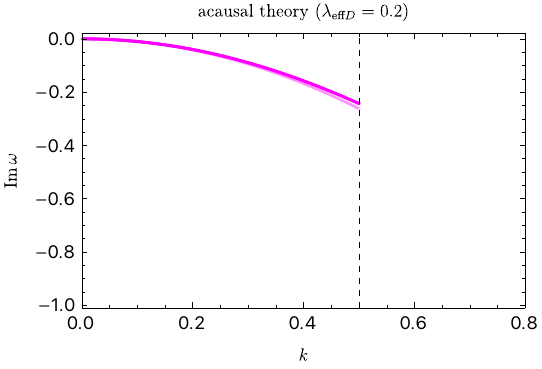}\,\,\includegraphics[width=0.4\textwidth]{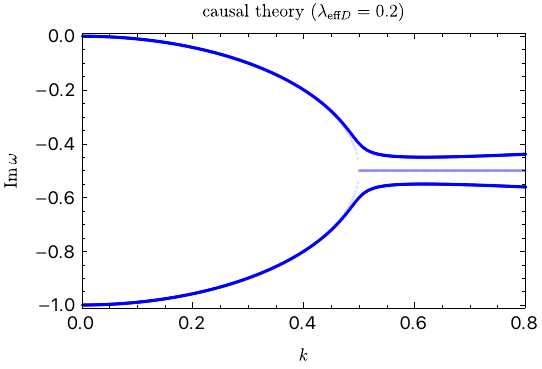} \\
	\caption{Pole-splitting due to the self-interactions. 
    In all panels, the linear-response dispersion relations, previously illustrated in Fig.~\ref{spectrum_linear}, are shown with low-opacity curves. 
    The dashed vertical line in the left panel shows the scale of emergence of new physics, beyond which the diffusion does not describe the whole physics. 
    \textbf{Top panel}: In both theories, the real parts of the modes separate into two distinct branches. 
    This leads to appearance of two and four modes in Theory A and Theory B, respectively. 
    \textbf{Bottom panel}: The imaginary parts of the modes in both theories are shifted due to the loop correction.  
    In Theory B, the four modes produced repel each other and no degeneracy occurs. 
    This is in contrast to the linear theory where the two modes merge at $k=0.5$. Note that there are four modes in Theory B, because each imaginary part in the bottom panel is associated with each real part shown in the top panel, giving four possible combination, see Fig.~\ref{split} for a representation of the modes in the complex-frequency plane. }  
	\label{int_modes}
\end{figure}

\begin{SCfigure}
	\centering
	\centering
	\includegraphics[width=0.5\textwidth]{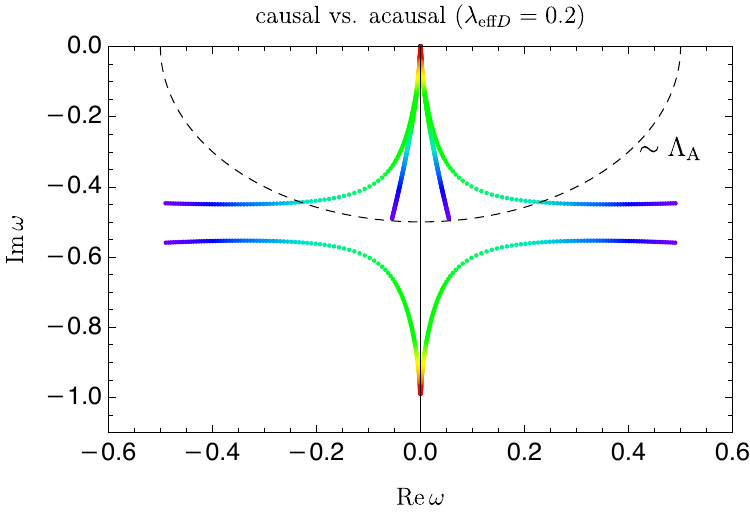}
	\caption{Pole splitting in Theory A and Theory B, in the complex-frequency plane. The two inner curves enclosed by the dashed semi-circle correspond to the dispersion relations of Theory A, while the four outer curves represent the four dispersion relations of Theory B. Each colored trajectory illustrates the evolution of the complex frequency $\omega$ of a single mode as the momentum 
$k$ varies from 0 (red) to $k_{max}$ (purple). We have used 
$k_{max}=0.5$ for Theory A and $k_{max}=0.8$ for Theory B. The naive cutoff of Theory A is shown by $\Lambda_{\text{A}}$.}
	\label{split}
\end{SCfigure}

As the first application, let us study the effect of loop corrections on the spectrum of Theory A and B. 
The spectrum, or the pole structure of the response function, can be obtained by solving the equation $G_{R}(\omega,\mathbf{k})^{-1}=0$. 
At the linear level, one just needs to solve $\big(G^{(0)}_{R}\big)^{-1}=0$. 
The result was already shown in Fig.~\ref{spectrum_linear}.
At the nonlinear level, this result is expected to get corrected. 
The leading correction comes from the cubic Lagrangian through one-loop effects: $\big(G^{(0)}_{R}+\delta G_R\big)^{-1}=0$. 
This equation can be written in terms of $\delta D$ as
\begin{equation}\label{spectral}
-i \tau \omega^2+\omega+i D k^2+ i \delta D \,k^2=0\;,
\end{equation}  
with $\delta D$ given by Eqs.~(9-A) and (9-B) in Table~\ref{Table}.
For Theory A, we should set $\tau=0$, as before.
The result of numerically solving Eq.~\eqref{spectral} is shown in Fig.~\ref{int_modes}.
For the sake of convenience we have defined $\lambda_{\text{eff} D}^2\equiv\frac{\lambda_D^2 T \chi}{16 D^{1/2}}$. 
The main effect of loop corrections on the spectrum is to split the poles; each of the poles is split into two poles. 
This can be seen more explicitly in Fig.~\ref{split}.

Notably, causality is preserved even at the nonlinear level. 
As shown in the right panel of Fig.~\ref{int_modes}, at large $k$, the four modes asymptotically approach the linear-response modes (see also Fig.~\ref{split}). The linear-response modes, by construction, describe a causal dynamics.

\subsection{Dynamic susceptibility and ``sum rules" }
The retarded Green's function or ``response function'' given by Eqs.~(8-A) and (8-B) is also called \textit{dynamic susceptibility}. 
It is convenient to write it as  
\begin{equation}\label{G_R_chi}
	G_{R}(\omega,\textbf{k})\equiv \chi(\omega , \textbf{k})=\chi'(\omega,\textbf{k})+i \chi''(\omega,\textbf{k})\;.
\end{equation}
Clearly, in both Theory A and Theory B, since $\sigma \equiv \chi D$ at $\omega=0$, we have $\chi(\omega=0 , \textbf{k})=\chi$, the thermodynamic susceptibility. 
Here $\chi'(\omega,\textbf{k})$ and $\chi''(\omega,\textbf{k})$ are \textit{reactive} and \textit{absorptive} parts of the retarded Green's function, respectively. 
In real space $\chi$ is a real-valued quantity. 
Thus $\chi^*(\omega,\textbf{k})=\chi(-\omega,-\textbf{k})$. 
Given this, it is easy to show that $\chi'(\omega,\textbf{k})$ and $\chi''(\omega,\textbf{k})$ are even and odd functions of $\omega$, respectively. 

Let us recall that $G_{R}(t,x)$ is defined for $t>0$. 
This causality condition is reflected in Fourier space, where $G_R(\omega, k)$ is analytic in the upper half-plane of the complex $\omega$-plane. 
Consequently, this property leads to the Kramers-Kronig dispersion relation \cite{Landau_1}:
 \begin{equation}\label{chi_decomp}
 	\chi(\omega,\textbf{k})=\int\frac{\mathrm{d}\omega'}{\pi}\frac{\chi''(\omega',\textbf{k})}{\omega'-\omega-i \epsilon}\;.
  \end{equation}
 This is a direct consequence of causality
 and tells us that if we know the absorptive (dissipative) part of the response function, we will have the whole response function. 
 This equation has also two implications which are commonly referred to as \textit{sum rules}:
 \begin{enumerate}
 	\item Setting $\omega=0$, we find 
 	\begin{equation}\label{sum_rule_1}\boxed{
 			\chi(\omega=0,\textbf{k})=\int_{-\infty}^{+\infty}\frac{\mathrm{d}\omega'}{\pi}\frac{\chi''(\omega',\textbf{k})}{\omega'}} \;.
 	\end{equation}
 	\item At $\omega\rightarrow +\infty$ we find 
 \begin{equation}\label{sum_rule_2}\boxed{
 	\lim_{\omega\rightarrow +\infty}	\omega^2\chi(\omega,\textbf{k})=-2\int_{0}^{+\infty}\frac{\mathrm{d}\omega'}{\pi}\omega'\,\chi''(\omega',\textbf{k})} \;,
 \end{equation}
 where we have used the fact that $\chi''(\omega,\textbf{k})$ is an odd function of $\omega$. 
 	\end{enumerate}
 Using $G_R$ from Eqs.~(8-A) and (8-B) in Table~\ref{Table}, and for the sake of simplicity considering the linear-response regime, i.e., $\delta D(\omega,\textbf{k}) = \delta \tau(\omega, \textbf{k}) =0$, it is straightforward to verify that the sum rule \eqref{sum_rule_1} holds for both Theory A and Theory B. 
 However, the sum rule \eqref{sum_rule_2} only holds for Theory B, where both sides evaluate to $-\chi D k^2/\tau$.
 For Theory A, both left- and right-hand sides diverge. 
 This indicates the acausal nature of Theory A, as these sum rules are direct consequences of causality and must hold in any causal theory of diffusion. We have checked that the same results hold at the nonlinear level.  
 
 This discrepancy can also be understood from a different perspective. 
 For Theory B, the integral in Eq.~\eqref{sum_rule_2} is inversely proportional to the relaxation time parameter. 
 On the other hand, Theory A is acausal because of it is the $\tau \rightarrow 0$ limit of Theory B. 
 Thus, the divergence of the integral is a direct consequence of the acausality of Theory A.

\begin{figure} [tb]
	\centering
	\centering
	\includegraphics[width=0.4\textwidth]{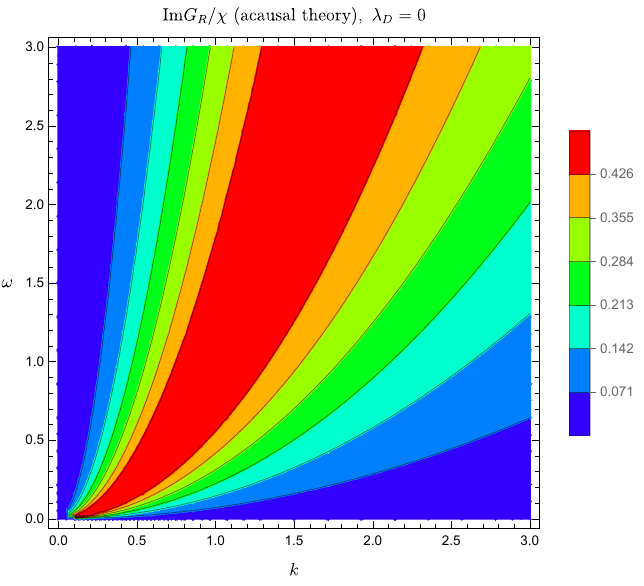} \,\,\,	\includegraphics[width=0.4\textwidth]{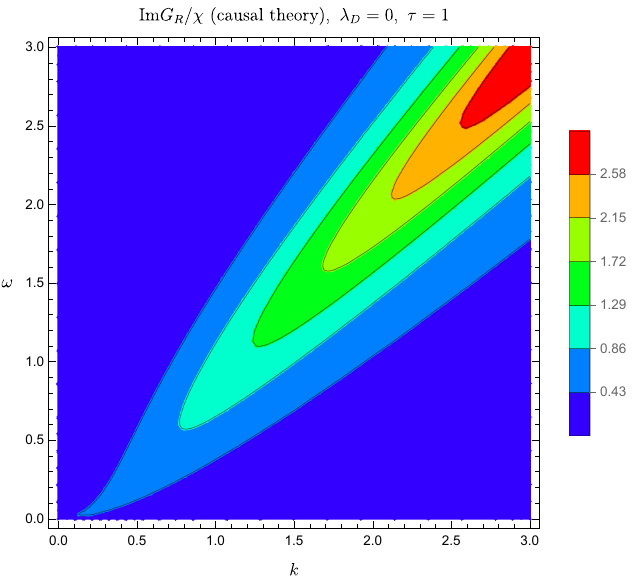}
	\includegraphics[width=0.4\textwidth]{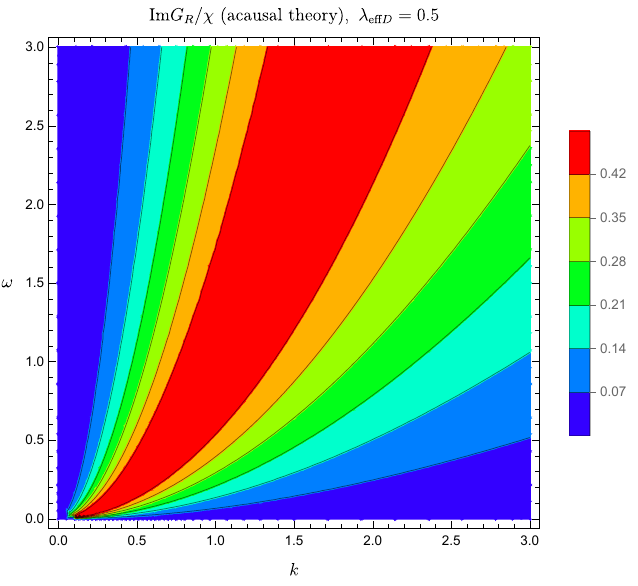} \,\,\,	\includegraphics[width=0.4\textwidth]{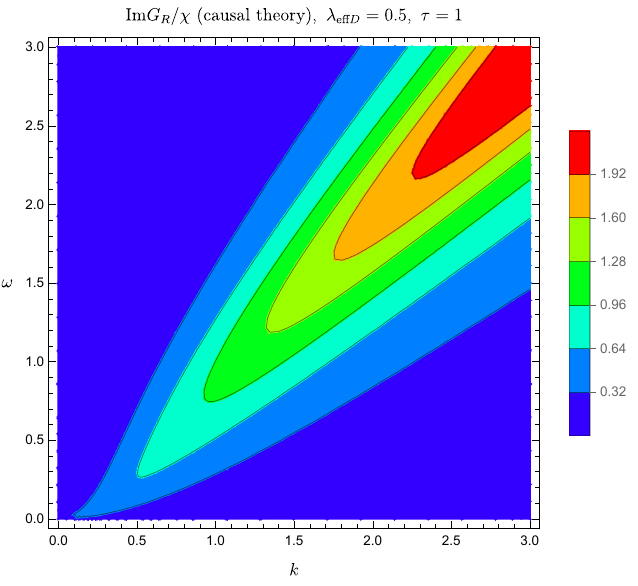}  \\
	\caption{Contour plot of  $\frac{\chi''(\omega,\textbf{k})}{\chi}$ as a function of $\omega$ and $k$. \textbf{Top row}:  Theory A (left) vs.~Theory B (right), both at linear response. \textbf{Bottom panel}: Theory A (left) vs.~Theory B (right) including one-loop corrections.  }
	\label{contour_plot}
\end{figure}
\begin{SCfigure}
	\centering
	\includegraphics[width=0.45\textwidth]{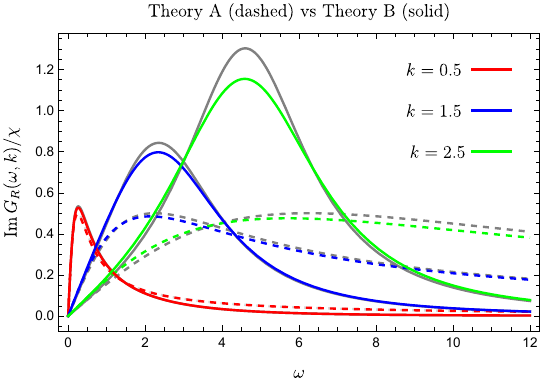}	
	\caption{The gray curves are the predictions of the free theory (for $k=0.5$, the gray curves are hidden by the red curves due to minimal corrections at small $k$.). The colored curves demonstrate how loop corrections influence the dynamic susceptibility. For the loop-corrected curves, we set $\lambda_{\text{eff}D}=0.5$.  For all solid curves, $\tau=0.25$ was used.\\
		}  
		\label{Sucseptibility}
	\end{SCfigure}

With this discussion in mind, we now proceed to compare $\chi''(\omega,k)$ between Theories A and B. The results are presented in Fig.~\ref{contour_plot}. 
Note that the plot in the top right panel was already given in Ref.~\cite{Glorioso:2022poi}. 
 Here are some comments:
\begin{itemize}
	\item[(a)] The main difference between Theory A and Theory B lies in the peak value of $\chi''(\omega,\textbf{k})$. For Theory A, in the linear-response regime, the peak is located at $\omega_{\text{max}} = Dk^2$ and its value remains constant, $\chi''_A(Dk^2,\textbf{k}) = \chi$ for any value of $k$. In contrast, for Theory B, the peak value increases with $k$, and this behavior is observed both with and without interactions. 
	\item[(b)] The effect of interactions (at any $k$) results in a broader and lower peak.
	\end{itemize}
To better understand these two observations and their justification, Fig.~\ref{Sucseptibility} illustrates $\text{Im} G_R \equiv \chi''$, for three distinct values of $k$. 
Both comments (a) and (b) can be explicitly observed in this figure.

Regarding (a), we should recall that the sum rule \eqref{sum_rule_1} applies to both Theory A and Theory B. 
This ensures that the integral under the curve (weighted by $1/\omega$) is constant. 
Starting with the dashed curve for $k=0.5$, the curve is non-vanishing at small values of $\omega$. 
For $k=1.5$, the characteristic frequency $\omega \sim D k^2$
increases, leading to a broader extent of the curve, as observed in the figure. 
Notably, this does not violate the sum rule \eqref{sum_rule_1}; at high frequencies, the value of the blue curve is weighted by a small $1/\omega$ factor. 
The integral of the additional contributions from the high-frequency domain compensates for the reduction at  small frequencies. 
Hence, a larger $k$ corresponds to broader peaks.

Besides simple mathematics, there is also a physical reason why the peak in all dashed gray curves has the same value. 
The key point is that $\chi''(\omega,\textbf{k})$ is proportional to the energy absorbed by the system during a disturbance, which eventually dissipates as heat. 
In Theory A, the rate of dissipation is $Dk^2$, while the diffusion time in this theory is simply $1/(Dk^2)$. 
As a result, one naturally expects a momentum-independent characteristic value for the total heat produced in the system. 
This characteristic value corresponds to the peak value of $\chi''(\omega,\textbf{k})$, which is indeed independent of $k$ and (at least in the linear-response regime) is equal to $\chi/2$, as illustrated in Fig.~\ref{Sucseptibility}.

To understand why, for the same value of $k$, the peak value in Theory B is larger than that in Theory A, recall that Theory B is a causal theory. 
As a result, its response function must decay more rapidly at high frequencies compared to Theory A, as is evident in the figure. 
Consequently, the sum rule \eqref{sum_rule_1} implies that, in order to maintain the same integral under the curve, the peak value in Theory B must be larger.

Regarding (b), it is evident that the primary effect of interactions is the coupling between different $k$-modes in Fourier space. 
This coupling broadens the peak, and to satisfy the sum rule, the peak value necessarily decreases. 
These features are clearly observed in Fig.~\ref{Sucseptibility}.
In summary, the main differences in the behavior of the dynamic susceptibility between Theory A and Theory B, as depicted in Fig.~\ref{Sucseptibility}, can be fully explained based on the sum rule or, equivalently, through arguments rooted in causality.

\subsection{Analytic structure of the correlation function in momentum space}
The correlation functions associated with the non-interacting part of the action is given by Eqs.~(5-A) and (5-B) in Table~\ref{Table}, respectively. 
Performing one-loop computations, one can find the correction to these expressions, see Appendix~\ref{loop} for details. 
It is found that 
\begin{align}\label{G_nn_A_corr}
	\delta G_{{nn}_{A}}(\omega,\mathbf{k}) &=
	\frac{\lambda_D^2 T^2 \chi^2}{4D^{1/2}}\,  k^4 \, 
	\frac{( D k^2 - 2i \omega)^{-\frac{1}{2}}}{( Dk^2 - i  \omega)^2} \;,
\\\label{G_nn_B_corr}
	\delta G_{{nn}_{B}}(\omega,\mathbf{k}) &=
	\frac{\lambda_D^2 T^2 \chi^2}{16D^{1/2}}\,  k^4 
	 \frac{[2 +  \tau D k^2 - 3i \tau \omega -(\tau \omega)^2]^2}{[ 1+\tau D k^2 -2i \tau \omega - (\tau \omega)^2]^{3/2}}  \frac{(D k^2 - 2i\omega  - \tau  \omega^2)^{-\frac{1}{2}}}{(Dk^2 - i \omega - \tau \omega^2)^2}\;.
\end{align}
 The squared expression in the denominator gives the free theory poles, i.e., Eqs.~(3-A) and (3-B) in Table~\ref{Table}, respectively. 
 The square-root expression, however, gives the branch-point singularities produced due to the interactions (see Eqs.~(11-A) and (11-B) in Table~\ref{Table}). 
 This can be understood as follows. 
 Let us consider the following loop diagram contributing to the self-energy, and consequently to $\delta D$:
 \begin{equation}\label{simple_loop}
 	\begin{split}
 		\scalebox{0.53}{	\begin{tikzpicture}[baseline=(a.base)]
 				\begin{feynman}
 					\vertex (a) ;
 					\vertex [right=of a] (a1) ;
 					\vertex [right=of a1] (a2);
 					\vertex [above right=of a2] (a3);
 					\vertex [below right=of a2] (a4);
 					\vertex [above right=of a4] (a5); 
 					\vertex [ right=of a5] (a6);
 					\vertex [ right=of a6] (a7);
 					\diagram* {
 						(a) --[very thick,->](a1)-- [boson,very thick] (a2) -- [quarter left, very thick,->] (a3)--[quarter left,boson,very thick](a5)--[quarter left,boson,very thick](a4)--[<-,quarter left, very thick](a2) ,
 						(a5) --  [very thick,->](a6)--  [boson,very thick](a7),
 					};\,.
 				\end{feynman}
				\node[] at (1.4,.4) {$\textbf{k}$};
				\node[] at (4,1.6) {$\textbf{k}'$};	
				\node[] at (4,-1.6) {$\textbf{k}-\textbf{k}'$};	
				\node[] at (6.6,.4) {$\textbf{k}$};					
 		\end{tikzpicture}} \, 
 	\end{split}
 \end{equation}
 where the solid and wavy lines correspond to the $n$ and $n_a$ fields in the Lagrangian, respectively. 
 For a given pair $(\omega,\textbf{k})$, any $\textbf{k}'$ that causes an internal line to go on-shell introduces a discontinuity in the loop integral. 
 This discontinuity can typically be resolved by deforming the integration contour. 
 However, if the integration contour becomes pinched by the singularities of two internal lines, the situation leads to what is known as a  ``\textit{pinch singularity}". 
 In such cases, the singularity cannot be avoided by any contour deformation. 
 As a result, the integral becomes singular at the pinch point. 
 Therefore, a pinch singularity in the integrand corresponds to a branch-point singularity in the loop integral.\footnote{See Refs.~\cite{Landau:1959, Eden:1966} for a detailed discussion in the context of quantum field theory (see also Refs.~\cite{Bourjaily:2020wvq,Huber:2023uzd}). Additionally, Section~4.2.2 of Ref.~\cite{Abbasi:2022rum} explores similar ideas within the framework of hydrodynamics, via solving the ``Landau loop equations".} 
\begin{figure}
	\centering
	\centering
	\includegraphics[width=0.49\textwidth]{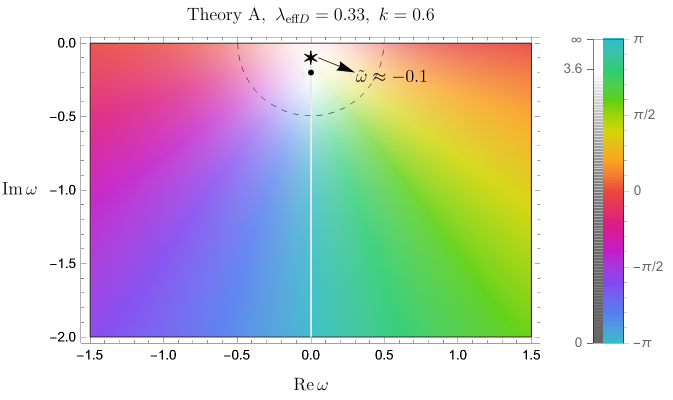}\includegraphics[width=0.49\textwidth]{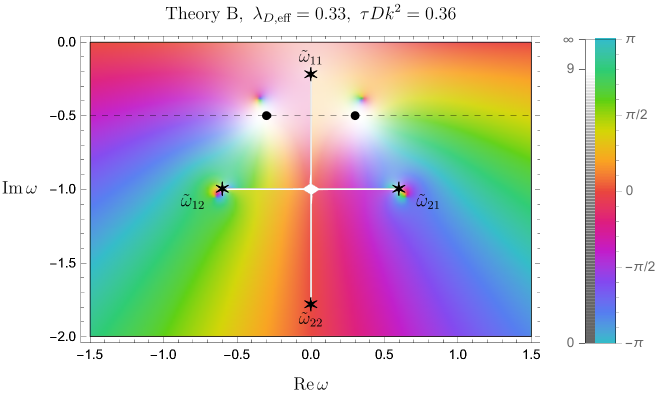} \\
		\includegraphics[width=0.49\textwidth]{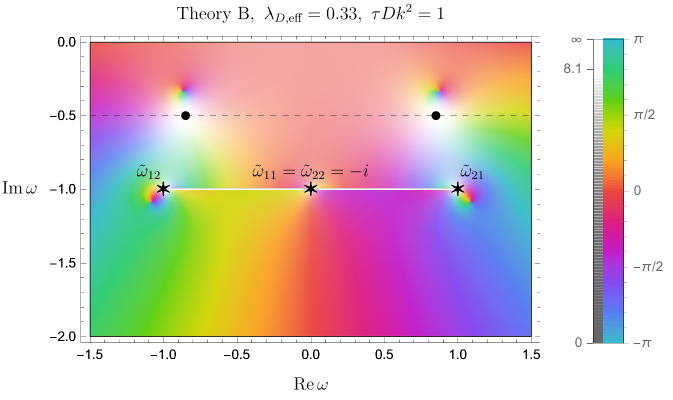}\includegraphics[width=0.49\textwidth]{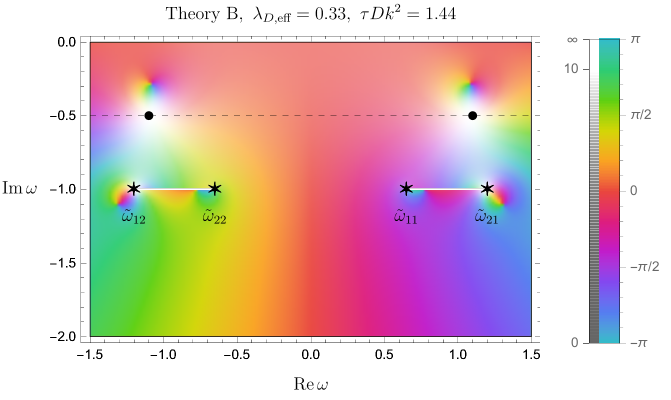} \\
	\caption{Analytic structure of $G_{nn}(\omega,\mathbf{k})=G^{(0)}_{nn}(\omega,\mathbf{k})+\delta G_{nn}(\omega,\mathbf{k})$. The color coding represents the phase of the complex function $G_{nn}(\omega,\mathbf{k})$. White lines indicate branch cuts, while black dots and stars mark the isolated poles and branch points, respectively. For clarity, these white dots (produced by Mathematica) are highlighted in black. \textbf{Top left:} In Theory A, there is only a single branch point.  \textbf{Top right:} For $\tau D k^2<1$ in Theory B, there are four branch points. As $\tau D k^2$ increases, the branch points $\tilde{\omega}_{11}$ and $\tilde{\omega}_{22}$ move toward each other, while $\tilde{\omega}_{12}$ and $\tilde{\omega}_{21}$ move away from each other.
    \textbf{Bottom left:} At $\tau D k^2=1$, $\tilde{\omega}_{11}$ and $\tilde{\omega}_{22}$ collide, reducing the discontinuity to a single horizontal branch cut. \textbf{Bottom right:} For $\tau D k^2>1$, $\tilde{\omega}_{11}$ and $\tilde{\omega}_{22}$ separate again, forming two distinct branch cuts. As $\tau D k^2$ increases further, each branch cut shrinks. In the limit $\tau D k^2\gg1$, the two branch cuts reduce to two second-order pole singularities. This behavior will be discussed in Fig.~\ref{contour_long_time_tail}.}  
	\label{analytic}
\end{figure}

To identify the pinch singularity, we first need to determine the singular points of the integrand, i.e., the values of $k'$ at which the internal lines go on-shell.  
For Theory B, there are four scenarios in which this might occur:
 \begin{eqnarray}\label{loop_up}
 	\omega'&=&\, \omega_{1,2}(\textbf{k}')\;,\\\label{loop_down}
 	\omega-\omega'&=&\,\omega_{1,2}(\textbf{k}-\textbf{k}')\;.
 \end{eqnarray}
 where, as before, 1 and 2 refer to the on-shell excitations of the free theory, given by Eq.~(3-B) in Table~\ref{Table}. 
 This leads to 
 \begin{equation}\label{combination_on_shell}
 	\omega=\,\omega_{1,2}(\textbf{k}')+\omega_{1,2}(\textbf{k}-\textbf{k}')\;.
 \end{equation}
   For each fixed value of $\textbf{k}'$, the above relation determines the full set of $(\omega,\textbf{k})$ pairs for which the integrand becomes singular. 
   To understand the analytic structure of the integral in the complex $\omega$-plane, it is instructive to examine how $\omega$ varies as a function of $\textbf{k}'$, while keeping the external momentum $k$ fixed. The corresponding values of $\omega$ trace out a branch cut in the complex $\omega$-plane, associated with the discontinuity of the loop integral.

This branch cut, in general, terminates at specific branch points. 
These endpoints can then be identified by extremizing $\omega(\textbf{k}')$, i.e., by solving for the stationary points of $\omega$ with respect to $\mathbf{k}'$. 
Therefore, these extrema correspond to threshold singularities, which physically represent the onset of two on-shell excitations within the loop in Eq.~\eqref{simple_loop}. 
Equivalently, they signify the locations at which the two singularities described in Eqs.~\eqref{loop_up} and \eqref{loop_down} pinch the integration contour.

This analysis must be carried out independently for each of the four channels: 
$(i,j)\in\{(1,1),(1,2),(2,1),(2,2)\}$ (see Appendix~\ref{singularities}). 
The resulting branch points, denoted by $\tilde{\omega}_{ij}$, align precisely with the square-root singularities derived via explicit loop calculations, as summarized in Eq.~(11-B) of Table \ref{Table}.

In contrast, Theory A possesses only a single excitation, as indicated in Eq.~(3-A) in Table~\ref{Table}. 
Therefore, Eq.~\eqref{combination_on_shell} can only be satisfied if both internal lines carry this excitation. 
Consequently, Theory A has only one branch point, as specified in Eq.~(11-A).\footnote{In analogy with scalar quantum field theory, Eqs.~(11-A) and (11-B) in Table~\ref{Table} correspond to the $p^2=(2m)^2$ singularity while Eqs.~(3-A) and (3-B) correspond to the pole $p^2=m^2$.}
In Fig.~\ref{analytic}, we show the analytic structure of the correlation function corrected at one-loop order.

\subsection{Real-time correlation function in momentum space}
The analytic structure depicted in Fig.~\ref{analytic} plays a crucial role in the calculation of real-time correlation functions.
Specifically, to evaluate the inverse Fourier integral  $G_{nn}(t>0,\mathbf{k})=\int_{-\infty}^{+\infty} \frac{\mathrm{d}\omega}{2\pi}G_{nn}(\omega,\mathbf{k})e^{- i \omega t}$, we must close the $\omega$-contour in the lower half of the complex $\omega$-plane. 
This process requires a careful handling of the branch-cut discontinuities and the poles illustrated in Fig.~\ref{analytic}. 
In the subsequent subsections, we will calculate $G_{nn}(t>0,\mathbf{k})$ analytically in two specific limiting cases, followed by a numerical evaluation for a broader range of parameters. 
\subsubsection{Analytical results}
In Ref.~\cite{Michailidis:2023mkd}, the inverse Fourier integral of $\delta G_{{nn}}(\omega,\mathbf{k})$ was evaluated analytically for Theory A. 
For Theory B, as discussed in the previous subsection, the analytic structure of  $\delta G_{{nn}}(\omega,\mathbf{k})$  is significantly more intricate. 
However, we have been able to analytically perform the inverse Fourier integral in two specific limits:
\begin{enumerate}
	\item \underline{$\tau D k^2\sim \tau \omega \ll1$:} In this limit, which we refer to as the \textbf{overdamped limit}, Eq.~\eqref{G_nn_B_corr}  simplifies to Eq.~\eqref{G_nn_A_corr}, i.e., Theory A and Theory B give the same behavior. 
    As a result, $\delta G_{{nn}}(\omega,\mathbf{k})$ in this limit is identical to the expression derived in Ref.~\cite{Michailidis:2023mkd}. 
    The analytic structure and the corresponding integration contour are depicted in the left panel of Fig.~\ref{contour_long_time_tail}. 
    The outcome of the integration is given by (see \App{App_overdamped} for a detailed derivation)
\begin{equation}\label{real_A_B}
\delta G_{{nn}_{A/B}}(t,\mathbf{k})= \frac{T^2\chi^2 \lambda_D^2}{4D^2}  |k| 
\,F_c(\tilde{t}_1)\;,
\end{equation}
where $F_c$ is a universal scaling function defined as 
\begin{align}\label{scale_A}
\tilde{t}_1\equiv t D k^2: & \quad F_c(\tilde{t}_1)= \left[ 
e^{-\frac{1}{2}\tilde{t}_1} (1 + \tilde{t}_{1}) \,\text{Erfi} \left( \sqrt{\frac{\tilde{t}_1}{2}} \right) - \sqrt{\frac{2 \tilde{t}_1}{\pi}}
\right]e^{-\frac{1}{2}\tilde{t}_1}\;,
\end{align}
where $\text{Erfi}(z) \equiv\text{Erf}(iz)/i$. 
In this limit, Theories A and B yield identical predictions; in other words $\tau D k^2\sim \tau \omega \ll1$ corresponds to the domain $k\lesssim k^*$ shown in Fig.~\ref{New_physics}. 
\textbf{ Equation \eqref{real_A_B} represents the correction to the late-time limit, $\tau \ll t$, of the overdamped function \eqref{over_damped}.}
\item \underline{$\tau D k^2 \gg1$:} This, which we refer to as the \textbf{underdamped limit}, is specific to Theory B, where each of the two branch cuts in the bottom right panel of Fig.~\ref{analytic} reduces to a second-order pole. 
The resulting analytical structure in this limit is illustrated in the right panel of Fig.~\ref{contour_long_time_tail}. 
By calculating the residues of the four second-order poles shown in the figure, we can analytically perform the inverse Fourier integral. 
The result is given by (see \App{App_underdamped} for a detailed derivation)
\begin{figure}
	\centering
	\begin{tikzpicture}
		[decoration={markings,
		}
		]
		\draw[help lines,->] (-3.5,0) -- (3.5,0) coordinate (xaxis);
		\draw[help lines,->] (0,-.2) -- (0,1) coordinate (yaxis);
				\draw[help lines,white] (0,-3.5) -- (0,-3);
		\draw[draw,line width=1.2pt,->] (-3,0) -- (-1,0) coordinate;
		\draw[draw,line width=1.2pt,->] (-1,0) -- (1,0) coordinate;
		\draw[draw,line width=1.2pt] (1,0) -- (3,0) coordinate;
		\path[draw,line width=1.3pt,blue,postaction=decorate] (-3,0) arc (180:266:3);
		\path[draw,line width=1.2pt,blue,postaction=decorate,->] (-3,0) arc (180:240:3);
		\path[draw,line width=1.2pt,blue,postaction=decorate,->] (.2,-3) arc (-86:-45:3);
		\path[draw,line width=1.2pt,blue,postaction=decorate] (.2,-3) arc (-86:0:3);
		
		\draw[draw,blue,line width=1.2pt,->] (-.2,-3) -- (-.2,-1.5) coordinate;
		\draw[draw,blue,line width=1.2pt] (-.2,-1.5) -- (-.2,-.4) coordinate;
		\draw[draw,blue,line width=1.2pt,->] (.2,-.4) -- (.2,-1.5) coordinate;
		\draw[draw,blue,line width=1.2pt] (.2,-1.5) -- (.2,-3) coordinate;
		\path[draw,line width=1.2pt,blue,postaction=decorate] (.2,-.4) arc (0:180:.2);

		\node[] at (0,-3) {$\times$};
		\node[] at (0,-2.8) {$\times$};
		\node[] at (0,-2.6) {$\times$};
		\node[] at (0,-2.4) {$\times$};
		\node[] at (0,-2.2) {$\times$};
		\node[] at (0,-2.) {$\times$};
		\node[] at (0,-1.8) {$\times$};	
		\node[] at (0,-1.6) {$\times$};	
		\node[] at (0,-1.4) {$\times$};
		\node[] at (0,-1.2) {$\times$};
		\node[] at (0,-1.) {$\times$};
		\node[] at (0,-.8) {$\times$};	
		\node[] at (0,-.6) {$\times$};	
		\node[] at (0,-.4) {$\times$};	
		\node[above] at (xaxis) {$\text{Re}\, \omega$};
		\node[left] at (yaxis) {$\text{Im} \,\omega$};
		\node[] at (.5,-.4) {$ \tilde{\omega}$};
	\end{tikzpicture}
	\begin{tikzpicture}
		[decoration={markings,
		}
		]
		\draw[help lines,->] (-3.5,0) -- (3.5,0) coordinate (xaxis);
		\draw[help lines,->] (0,-3.5) -- (0,1) coordinate (yaxis);
		\draw[draw,line width=1.2pt,->] (-3,0) -- (-1,0) coordinate;
		\draw[draw,line width=1.2pt,->] (-1,0) -- (1,0) coordinate;
		\draw[draw,line width=1.2pt] (1,0) -- (3,0) coordinate;
		\path[draw,line width=1.3pt,blue,postaction=decorate] (-3,0) arc (180:270:3);
		\path[draw,line width=1.2pt,blue,postaction=decorate,->] (3,0) arc (0:-90:3);
		\path[draw,line width=1.2pt,blue,postaction=decorate] (0,-3) arc (-90:0:3);
		
		\node[] at (1.5,-1) {$\boldsymbol{\bullet}$};
		\node[] at (-1.8,-1) {$\boldsymbol{\bullet}$};
		
		\node[] at (1.5,-2) {$\boldsymbol{\times}$};
		\node[] at (1.5,-2) {$\boldsymbol{+}$};
		\node[] at (-1.8,-2) {$\boldsymbol{\times}$};
		\node[] at (-1.8,-2) {$\boldsymbol{+}$};
		
		\node[above] at (xaxis) {$\text{Re}\, \omega$};
		\node[left] at (yaxis) {$\text{Im} \,\omega$};

		\node[scale=0.6] at (1.6,-.6){$\frac{-i}{2\tau}+\sqrt{\tau D k^2}$};
		\node[scale=0.6] at (-1.6,-.6){$\frac{-i}{2\tau}-\sqrt{\tau D k^2}$};
		
		\node[scale=0.6] at (1.6,-1.6){$\frac{-i}{\tau}+\sqrt{\tau D k^2}$};
		\node[scale=0.6] at (-1.6,-1.6){$\frac{-i}{\tau}-\sqrt{\tau D k^2}$};
	\end{tikzpicture}
	\caption{\textbf{Left:} In the overdamped limit  $\tau D k^2\sim \tau \omega\ll1$, there is only one branch cut, originating from $\tilde{\omega}=-\frac{i}{2}Dk^2$ and extending to negative imaginary infinity. This corresponds to the region enclosed by the dashed semicircle in the top left panel of Fig.~\ref{analytic}. The integration contour is deformed for this case, ensuring that the branch cut is properly encompassed and its contribution to the inverse Fourier integral is accounted for.  \textbf{Right:} In the underdamped limit $\tau D k^2 \gg1$, the two branch cuts, illustrated in the bottom right panel  of Fig.~\ref{analytic}, contract into two second-order poles, marked by stars. This occurs because in this limiting scenario, $\tilde{\omega}_{11}\rightarrow \tilde{\omega}_{21}$ and $\tilde{\omega}_{22}\rightarrow \tilde{\omega}_{12}$.  The contour is closed in the lower half-plane.  The two dots also are second-order poles; see the black dots in the bottom right panel of Fig.~\ref{analytic}.}
	\label{contour_long_time_tail}
\end{figure}
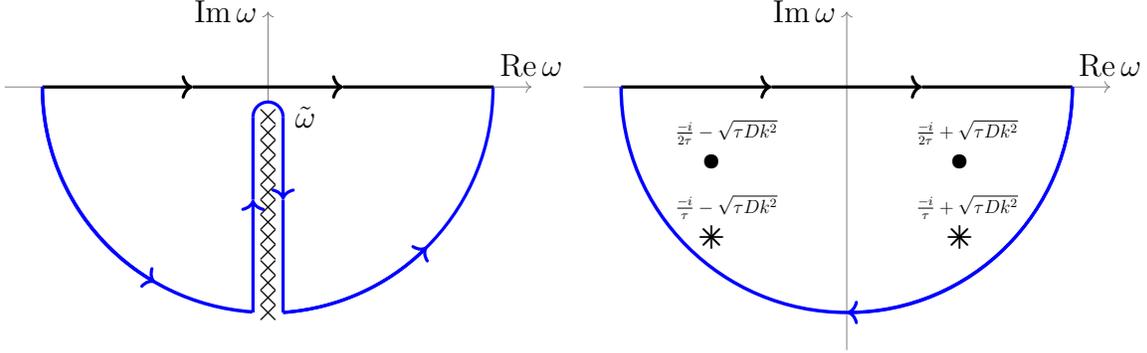
%
\begin{equation}\label{real_B}
\begin{split}
\delta G_{{nn}_{B}}(t,\mathbf{k})=\frac{T^2 \chi^2 \lambda_D^2}{4D^2}\frac{\tau D k^2}{\sqrt{\tau D}}\,G_c(\tilde{t}_2)\;,
\end{split}
\end{equation}
where $G_c$ is a universal scaling function defined as 
\begin{align}\label{scale_B}
\tilde{t}_2\equiv \frac{t}{\tau}: &\quad G_c(\tilde{t}_2)= \Bigg[\left(1-\frac{\tilde{t}_2}{2}\right)e^{-\frac{\tilde{t}_2}{2}}-\left(1+\frac{\tilde{t}_2}{8}\right)e^{-\tilde{t}_2}\Bigg]\cos\left(f_0\tilde{t}_2\right)\\\nonumber
&\,\,\,\,\,\,\,\,\,\,\,\,\,\,\,\,\,\,\,\quad+ \Bigg[\left(1-\frac{\tilde{t}_2}{4}\right)e^{-\frac{\tilde{t}_2}{2}}-\frac{1}{8}\left(7+\tilde{t}_2\right)e^{-\tilde{t}_2}\Bigg]\frac{\sin\left(f_0\tilde{t}_2\right)}{f_0}\;,
\end{align}
with $f_0=\sqrt{\tau D k^2}$.

\textbf{Equation \eqref{real_B} is in fact the correction to the  high-frequency limit, $\tau D k^2\gg1$, of the underdamped function \eqref{under_damped}.}
\end{enumerate}
\begin{figure}[tb!]
	\centering
	\centering
	\includegraphics[width=0.45\textwidth]{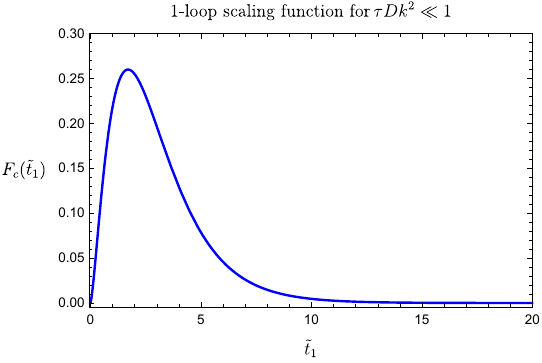}\,\,\,\,\,\,\includegraphics[width=0.45\textwidth]{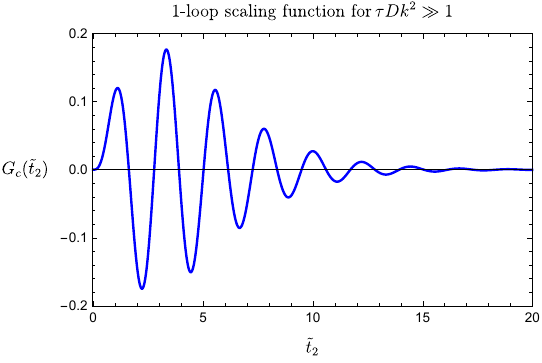} \\
	\caption{Scaling functions. \textbf{Left:} The function $F_c$ determines the universal part of the one-loop correction to the overdamped real-time correlation function \eqref{over_damped}, as a function of $\tilde{t}_1=t D k^2$.  \textbf{Right:} The function $G_c$ determines the universal part of the one-loop correction to the underdamped real-time correlation function \eqref{under_damped}, as a function of $\tilde{t}_2=t/\tau$ and the frequency parameter $f_0=\sqrt{\tau D k^2}$. Here we have set $f_0=\sqrt{8}$.}  
	\label{scaling_functions}
\end{figure}

Interestingly, we see that in the above two limits,  the real-time dynamics of $\delta G_{{nn}}(t,\mathbf{k})$ is given in terms of two scaling functions. 
While Eq.~\eqref{scale_A} is a universal function of $\tilde{t}_1=tDk^2$, Eq.~\eqref{scale_B} is a universal function of $\tilde{t}_2=t/\tau$ modulo the value of the large parameter $f_0\gg1$.

The universality of these functions is a direct prediction of the SK-EFT. 
Once $\lambda_D=\mathrm{d} D(n)/\mathrm{d}n$  is known,\footnote{This can be done by measuring the diffusivity at several densities \cite{Michailidis:2023mkd}.} the entire function $\delta G_{nn}(t,\mathbf{k})$ is fixed. 
While the correction in Eq.~\eqref{real_A_B} provides insight into intermediate-time dynamics \cite{Michailidis:2023mkd}, Eq.~\eqref{real_B} offers valuable information about the times of the order of relaxation time $\tau$. 
This distinction highlights a key difference between the predictions of Theory A, as presented in Ref.~\cite{Michailidis:2023mkd}, and those of Theory B found in this work.
In Fig.~\ref{scaling_functions}, we have illustrated the two scaling functions.

\subsubsection{Numerical results}
\begin{figure} [htbp] 
	\centering
	\centering
	\includegraphics[width=0.33\textwidth]{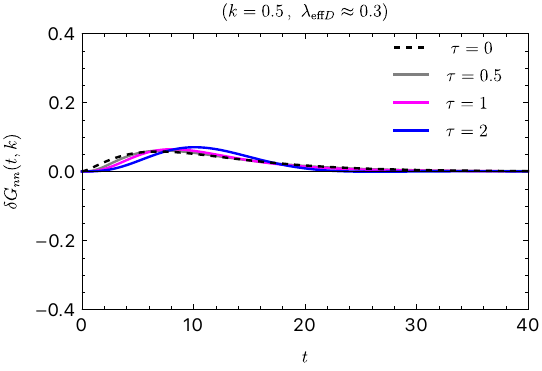}\includegraphics[width=0.33\textwidth]{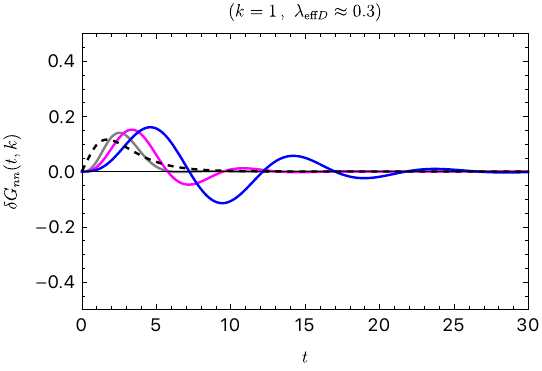}\includegraphics[width=0.33\textwidth]{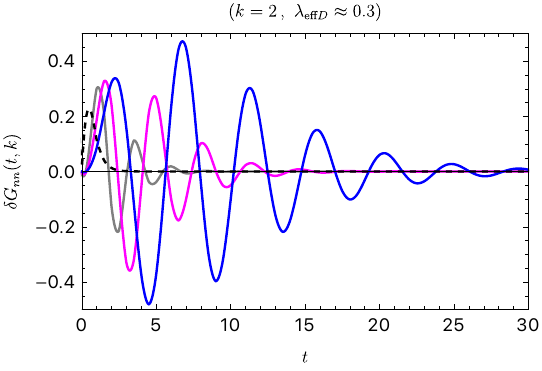} \\
	\caption{Numerical computation of the inverse Fourier integral. As before $\tau=0$ corresponds to Theory A; it is seen that the dashed curves matches with the scaling function $F_c$ in Fig.~\ref{scaling_functions}. The blue curve in the right panel corresponds to $f_0=\sqrt{8}$  in Fig.~\ref{scaling_functions}. This curve exhibits a perfect match with the scaling function $G_c$.}  
	\label{branch_cut}
\end{figure}

We have also numerically evaluated the inverse Fourier integral for both Theory A and Theory B, covering a broad range of $\tau D k^2$. 
The results are presented in Fig.~\ref{branch_cut}, where we show them for three different values of $k$, each calculated for four distinct values of $\tau$. 
This way we explore 12 different situations, ranging from very small to relatively  large values of $\tau D k^2$. 
Considering the parameter used in the figure, a perfect agreement with the scaling functions of  Fig.~\ref{scaling_functions} is seen. 
All dashed curves, corresponding to Theory A, match with $F_c$, while the blue curve in the right panel matches with $G_c$.

\section{Discussion and outlook}
\label{discussion}
In this work, we analyzed Theory A, which represents acausal diffusion, and Theory B, which describes causal diffusion with two modes well separated from the rest of the non-hydrodynamic spectrum.
We demonstrated that Theory B exhibits two analytically controlled limits: the overdamped and underdamped limits. 
Notably, we show that the overdamped limit reproduces the predictions of Theory A with high precision.

Theory A provides a universal description of the late-time dynamics of a single diffusive charge in all thermal systems. 
In contrast, Theory B is not universal in this sense; however, it applies to a broad class of physical systems, including cold-atom systems \cite{Brown:2019}, QCD near the critical point \cite{Du:2021zqz}, magnetohydrodynamics \cite{Dash:2022xkz}, and others.

More specifically, in this work we compared Theory A and Theory B within the framework of  Schwinger-Keldysh effective field theory (SK-EFT). 
Our comparison is based on the calculation of fluctuation observables within the SK-EFT, taking into account nonlinear interactions. 
In particular, we compute the one-loop correction to the real-time density correlation function, $G_{nn}(t,\mathbf{k})=G_{nn}^{(0)}(t,\mathbf{k})+\delta G_{nn}(t,\mathbf{k})$, in Theory B, focusing on two asymptotic limits mentioned above.

In the \textbf{overdamped limit}, $\tau D k^2\ll 1$, we recover results consistent with the acausal theory (Theory A) \cite{Chen-Lin:2018kfl}. 
The density-density correlation function
\begin{equation}\label{G_nn_B_1}
G_{nn}(t,\mathbf{k})= T \chi\,
	 F_0(\tilde{t}_1)+ \, \frac{T^2\chi^2 \lambda_D^2}{4D^2} |k| F_c	(\tilde{t}_1)\,,\,\,\,\,\,\,\,\,\,\,\tilde{t}_1=t D k^2\;,
\end{equation}
consists of two parts: The linear response of the system, arising from the non-interacting part of the EFT, is fully described by a universal scaling function $F_0(\tilde{t}_1)=e^{-\tilde{t}_1}$. 
The nonlinear response of the system, resulting from interactions in the EFT, is characterized by a universal dimensionless function $F_c(\tilde{t}_1)$ given by Eq.~\eqref{scale_A} and a non-universal factor.

As demonstrated in Ref.~\cite{Michailidis:2023mkd}, these results can serve as a precision test for diffusive systems. 
The following three lessons can be learned \cite{Michailidis:2023mkd}:
\begin{itemize}
\item  The non-universal factor in the second part can be fixed by linear-response data. 
This is because $\lambda_D=\mathrm{d}D(n)/\mathrm{d}n$, where $D$ is measured through the linear response of the system. 
This highlights the predictive power of EFT in diffusive systems; linear-response experiments can determine $D(n)$, enabling predictions for nonlinear responses.
\item While the linear-response part of $G_{nn}(t,\mathbf{k})$ accurately predicts late-time dynamics, the one-loop corrections in Eq.~\eqref{G_nn_B_1} extend the predictive power of EFT to intermediate times. 
\item Interestingly, interactions become irrelevant in the infrared (IR) limit, which ensures the validity of perturbation theory without the need for fine-tuning. 
This can be understood by examining the ratio of the coefficients of the second term to the first term in Eq.~\eqref{G_nn_B_1}.
Using $\lambda_D = \frac{\mathrm{d}D}{\mathrm{d}n} \sim \frac{D}{n}$ as well as $n \sim \chi \mu$, where $\mu$ is the chemical potential in equilibrium, we estimate
$$
\frac{1}{T \chi} \frac{T^2 \chi^2 \lambda_D^2}{4D^2}|k| \sim \frac{T \chi \frac{D^2}{n^2}}{D^2}\sqrt{k^2} \sim \frac{T \chi}{(\chi \mu)^2} \sqrt{k^2} = \frac{\chi_0}{\chi} \sqrt{\tau D k^2}\;,
$$
where $\chi_0\sim \frac{T}{\mu^2\sqrt{\tau D}}$.
The smallness of this ratio in the IR justifies the perturbative expansion. Moreover, including higher-order loop corrections can systematically improve the accuracy of the EFT predictions.
\end{itemize}
In the \textbf{underdamped limit}, $\tau D k^2\gg1$, we analytically derive 
\begin{equation}\label{G_nn_B_2}
G_{nn}(t,\mathbf{k})= T \chi\,
	 G_0(\tilde{t}_2)+ \, \frac{T^2\chi^2 \lambda_D^2}{4D^2} \frac{\tau D k^2}{\sqrt{\tau D}} G_c(\tilde{t}_2)\,,\,\,\,\,\,\,\,\,\,\,\tilde{t}_2=\frac{t}{\tau}\;,
\end{equation}
where $G_0(\tilde{t}_2)= \Big[\frac{1}{2f_0}\sin\big(f_0\tilde{t}_2 \big)+\cos\big(f_0 \tilde{t}_2\big)\Big]e^{-\frac{\tilde{t}_2}{2}}$ and $G_c$ is given by Eq.~\eqref{scale_B}. 
The ratio of the coefficients of the second term to the first term in Eq.~\eqref{G_nn_B_2}, which is of the order $\frac{\chi_0}{\chi}\,\tau D k^2$, can become large in the limit $\tau D k^2\gg1$. 
In such a regime, perturbation theory would no longer be controlled unless one introduces a form of fine-tuning. 
This can be achieved by taking the susceptibility $\chi$ to be sufficiently large, thereby suppressing the ratio and restoring the validity of the perturbative expansion. 
Hence, the EFT calculations presented here remain reliable for systems characterized by a large susceptibility.

Under this assumption, and similarly to the overdamped limit, one can use linear-response results to make predictions for the nonlinear response, at timescales of order $\tau$. 
Following the approach of Ref.~\cite{Michailidis:2023mkd}, these predictions could be tested through dedicated experiments or simulations. 
Such tests would serve as precision probes of Theory B in the underdamped regime, a task we leave for future work.

In this paper, we have used the Fourier-space results of Refs.~\cite{Chen-Lin:2018kfl} and \cite{Abbasi:2022aao}. 
Notably, the results from Ref.~\cite{Abbasi:2022aao} pertain to the special case where $\tau=$ const.\ and $\sigma=$ const.\ (see Appendix~\ref{loop} for further details). 
It would be interesting to explore how allowing $\tau$ and $\sigma$ to vary with density affects the real-time dynamics of the correlation functions studied in this work. 
We leave this extension to future investigation.

Previous works (e.g., Refs.~\cite{Chen-Lin:2018kfl,Michailidis:2023mkd}) largely focused on the overdamped limit $\tau D k^2\ll 1$, where the EFT matches acausal diffusion.
Our analysis extends this by demonstrating that the one-loop SK-EFT remains analytically tractable even in the underdamped or ``quasi-diffusive" limit $\tau D k^2 \gg 1$, revealing a richer analytic structure in real time.

The results presented here extend the predictive power of SK-EFT beyond the conventional diffusive regime by incorporating a physical relaxation time $\tau$, leading to a quasi-diffusive framework with analytically accessible scaling behavior. 
While the diffusion constant $D$ and relaxation time $\tau$ can be determined from low-frequency linear-response experiments, the non-universal pre-factor in the sub-leading component $G_c$	of the real-time correlator provides insight into the nonlinear regime. 
Once calibrated through nonlinear-response measurements, such as high-frequency density modulations or driven quenches, the coupling constant $\lambda_D$ enables the EFT to predict new observables, including three-point correlation functions.

This framework is particularly well-suited for simulation-based verification in fermionic systems, such as the ultra-cold gas of $^6\text{Li}$ in optical lattices studied in Ref.~\cite{Brown:2019}. 
Large-scale determinant Quantum Monte Carlo (DQMC) calculations \cite{Varney:2009} have enabled the observation of density correlations at the single-atom level, thereby facilitating detailed investigations of many-body states in optical lattice setups. 
These developments raise the possibility of testing the quasi-diffusive effective field theory developed here, particularly in the underdamped regime $\tau D k^2\gg 1$, and assessing the predicted universal structures in nonlinear correlation functions.

In a different direction, it would be very interesting to explore the ideas discussed in this paper within the context of holography.
At the level of linear response, the AdS gravity duals corresponding to Theory A and Theory B are already known. 
Theory A corresponds to linear fluctuations of a Maxwell field on a fixed AdS–Schwarzschild background \cite{Son:2007vk}. 
In Ref.~\cite{Grozdanov:2018fic}, it is shown that Theory B can be realized as fluctuations of the background itself in AdS when the bulk theory includes Einstein–Gauss–Bonnet gravity. 
It would be particularly interesting to understand how the nonlinear effects in Theory A and Theory B can be realized on the gravity side. 
We leave the investigation of this question to future work.


Another direction is to investigate fluctuation dynamics in systems without conserved charges, such as Brownian motion described by Langevin dynamics \cite{Oei:2024zyx, Son:2009vu, Casalderrey-Solana:2007ahi, Casalderrey-Solana:2006fio, Rajagopal:2025ukd, Rajagopal:2025rxr}. 
It would be worthwhile to explore how SK-EFT structures manifest in quark diffusion (see Refs.~\cite{Liu:2018kfw,Lin:2023bli} for related works) and whether universal scaling behaviors, similar to those identified in our quasi-diffusive regime, can be uncovered in such non-hydrodynamic settings.


Finally, we observe structural parallels between Theory B and Model-B critical dynamics \cite{Hohenberg:1977ym}, particularly in how the relaxation time modifies the spectrum of diffusive modes. 
Incorporating aspects of critical behavior into our framework, along the lines of Ref.~\cite{Du:2021zqz}, may enable systematic comparisons with results obtained from functional renormalization group approaches Ref.~\cite{Roth:2023wbp}, as well as with other real-time analyses developed in Refs.~\cite{Wu:2019qfz,Sakai:2025xmo}. 
Such comparisons could clarify the interplay between nonlinearities, scaling, and causality in near-critical diffusive systems.

	\section*{Acknowledgments}
	We thank Xin-Nian Wang for his collaboration during the early stages of this project. In particular, N.A. gratefully acknowledges the inspiring discussions during the Institute of Particle Physics (IOPP) colloquium at Central China Normal University (CCNU) Wuhan, which helped initiate this work, and he appreciates Hui Huang for early discussions related to numerical aspects of the work.
	N.A.\ would like to thank the Institute for Theoretical Physics of Goethe University for the warm hospitality during the completion of this work, where he was supported in part by the ExtreMe Matter Institute EMMI at the GSI Helmholtzzentrum für Schwerionenforschung, Darmstadt, Germany,  as well as by grant number 561119208 “Double First Class” start-up funding of Lanzhou University, China.
	D.H.R.\ acknowledges support by the Deutsche Forschungsgemeinschaft (DFG, German Research Foundation) through the CRC-TR 211 ``Strong-interaction matter under extreme conditions'' – project number 315477589 – TRR 211 and
	by the State of Hesse within the Research Cluster ELEMENTS (Project ID 500/10.006). {M.K.\ was supported in part by the U.S.~Department of Energy grant DE-SC0012447. All authors thank the ECT*, Trento, Italy and the Galileo Galilei Institute for Theoretical Physics for their hospitality and the INFN for partial support during the completion of this work.}

	\appendix
	\section{Correlation function from stochastic equations }
	\label{App_Landau}
	As discussed in Ref.~\cite{Landau_1}, one way to incorporate the effect of fluctuations in hydrodynamic equations is to promote them to stochastic equations. 
    This is done by adding stochastic noise terms to the constitutive relation of currents. Recalling Eqs.~(1-A) and (1-B) of Table~\ref{Table}, together with the conservation of density: 
	\begin{equation}\label{noise}
		\begin{split}
			\partial_t   \breve{n}(t,\textbf{k})  +i \textbf{k}\cdot \breve{\textbf{J}}(t,\textbf{k}) =&\,0\;,\\
			\tau\, \partial_t  \breve{\textbf{J}}(t,\textbf{k}) + \breve{\textbf{J}}(t,\textbf{k}) + i\,D \,\textbf{k} \,  \breve{n}(t,\textbf{k}) =&\,\xi(t,\textbf{k})\;,
		\end{split}
	\end{equation}
	 where $\xi$ is the random-noise current and $\breve{n}$ and $\breve{\textbf{J}}$ are the noisy density and current, respectively. 
    Then the correlation function of the noisy field $\breve{n}$ is defined as 
	\begin{equation}\label{def_G_nn}
	G_{nn}(t,\textbf{k}):=\langle \breve{n}(t,\textbf{k})\breve{n}(0,-\textbf{k})\rangle \;.
	\end{equation}
	Here, the averaging is over the values of $\breve{n}$ at the times $t$ and $0$. 

Each noisy field can be considered as the sum of two parts, an average and a fluctuating part, e.g., $\breve{n}= n + \delta \breve{n}$.\footnote{Note that in the purely dissipative picture, when the fluctuations are neglected, one only deals with the average field. The latter itself is considered to be given as $n=\bar{n}+\delta n$ near equilibrium, where $\bar{n}$ is the equilibrium solution to the dissipative equation of the average field $n$.} When the fluctuation is small compared to the average,  $G_{nn}$ in Eq.~\eqref{def_G_nn} can be written as
\begin{equation}\label{def_G_nn_no_noise}
	G_{nn}(t,\textbf{k}):=\langle n(t,\textbf{k})n(0,-\textbf{k})\rangle\;.
\end{equation}
There are a couple of comments regarding this expression:
\begin{itemize}
		\item	It is clear from the definition that, for systems which are invariant under time translations, $G_{nn}(t,\textbf{k})=G_{nn}(-t,-\textbf{k})$. This then enforces the spectral function $G_{nn}(\omega,\textbf{k})$ to be real-valued. 
	\item
	Note that the average follows the noise-averaged version of Eq.~\eqref{noise} for $t>0$:
	\begin{equation}\label{noise_ave}
		\begin{split}
			\partial_t   n(t,\textbf{k})  +i  \mathbf{k}\cdot\mathbf{J}(t,\textbf{k}) =&\,0\;,\\
			\tau\, \partial_t  \mathbf{J}(t,\textbf{k}) + \mathbf{J}(t,\textbf{k}) + i\,D \,\textbf{k} \,  n(t,\textbf{k}) =&\,0\;.
		\end{split}
	\end{equation}
	\item Considering Eq.~\eqref{noise_ave}, the average in Eq.~\eqref{def_G_nn_no_noise} can then be interpreted as the average over the probabilities of the various values of $n$ at $t=0$. Since we are interested in Eq.~\eqref{def_G_nn_no_noise} at late time, i.e., $t$ being much larger than the local equilibration time, we can consider $G_{nn}(t=0,\textbf{k})$  as being equal to its equilibrium value:
	\begin{equation}\label{bdy}
		\langle n(0,\textbf{k})n(0,-\textbf{k})\rangle =\,T \chi\;.
	\end{equation}
Then, in order to find $\langle n(t,\textbf{k})n(0,-\textbf{k})\rangle $, we only need to construct the equations governing dynamics of $\langle n(t,\textbf{k})n(0,-\textbf{k})\rangle$ and solve them with the boundary condition \eqref{bdy}. 
As was discussed in the text (see the discussion around Eq.~\eqref{boundary_condition}), we will need one more boundary condition which ensures causality.
\end{itemize}	
We start by multiplying Eq.~\eqref{noise_ave} with $n(0,-\mathbf{k)}$ and average as explained above. 
We find
	\begin{equation}\label{coupled_correlations}
		\begin{split}
			\partial_t \langle  n(t,\textbf{k})  n(0,-\textbf{k})\rangle+i k_i \langle J_i(t,\textbf{k})  n(0,-\textbf{k})\rangle=&\,0\;,\\
			\tau\, \partial_t \langle J_i(t,\textbf{k})  n(0,-\textbf{k})\rangle+\langle J_i(t,\textbf{k})  n(0,-\textbf{k})\rangle+ i\,D \,k_i \langle  n(t,\textbf{k})  n(0,-\textbf{k})\rangle=&\,0\;.
		\end{split}
	\end{equation}
Let us introduce the simple notation $G_{\phi_a\phi_b}(t,\textbf{k})\equiv\langle  \phi_a(t,\textbf{k})  \phi_b(0,-\textbf{k})\rangle$. 
Below Eq.~\eqref{def_G_nn_no_noise}, we pointed out the symmetry property of $G_{\phi_a\phi_b}(t,\textbf{k})$ for the case $\phi_a,\phi_b\equiv n$. 
This is true because in this case, $\phi_a$ and $\phi_b$ do not change sign under time reversal. 
If both $\phi_a$ and $\phi_b$ change sign under time reversal, then the same symmetry relation will be valid. 
However, if only one of the two changes sign, then we will get an extra minus sign. For instance, $G_{J_in}(t,\textbf{k})=-G_{J_in}(-t,-\textbf{k})$.

The main goal is to find the spectral function $G_{nn}(\omega,\mathbf{k})$. 
To solve Eqs.~\eqref{coupled_correlations}, we introduce the one-sided Fourier transform as follows 
\begin{equation}
G^{(+)}_{\phi_a \phi_b}(\omega,\textbf{k})=\,\int_0^{+\infty}\mathrm{d}t\, G_{\phi_a \phi_b}(t,\textbf{k}) e^{i\omega t}\;.
	\end{equation}
Now, we multiply Eqs.~\eqref{coupled_correlations} by $e^{i\omega t}$ and integrate over $0$ to $+\infty$. 
Using the boundary conditions \eqref{boundary_condition} and \eqref{boundary_condition_2} as well as the fact that the correlation functions vanish when $t \rightarrow \infty$, we find 
	\begin{equation}
		\begin{split}
			- T\chi -i \omega  \, G^{(+)}_{nn}(\omega,\textbf{k})+ik_i G^{(+)}_{J_in}(\omega,\textbf{k})=&\,0\;,\\
			(1-i \omega \tau) G^{(+)}_{J_in}(\omega,\textbf{k})+i D k_i G^{(+)}_{nn}(\omega,\textbf{k})=&\,0\;.
		\end{split}
	\end{equation}
	Solving for $G_{nn}^{(+)}$ we find 
	\begin{equation}\label{correlation}
		G^{(0)}_{nn}(\omega, \textbf{k})=\int_{-\infty}^{+\infty}\mathrm{d}t\, G_{nn}(t,\textbf{k}) e^{i\omega t}\equiv2 \text{Re} \,G_{nn}^{(+)}(\omega,\textbf{k})=\,\frac{2\,T\chi\,Dk^2}{\omega^2+(\tau \omega^2 -D k^2)^2}\;.
	\end{equation}
	This was given by Eq.~(5-B) in Table~\ref{Table}. 

    	Here we calculated the correlation function in momentum space. However, if only the response function is of interest, one can introduce a source term $\chi D \nabla \mu$ in the second line of Eq.~\eqref{noise_ave} and directly obtain $n(t,\mathbf{k})$.
        The response function can then be determined as the variation of $n(\omega,\textbf{k})$  with respect to $\mu(\omega, \textbf{k})$, see, e.g., Ref.~\cite{Chagnet:2023xsl}. 
	\section{Correlation function from SK-EFT}
	\label{corrfunc_SKEFT}
	The correlation function \eqref{correlation} can be also derived from SK-EFT. 
    One just needs to consistently couple the fields to background sources and then vary the effective action (more precisely the generating functional) with respect to the sources.  
    After an integration by parts, we can rewrite the quadratic Lagrangian (4-B) in the following form 
\begin{equation}
		\mathcal{L}_2=iT \sigma  \partial_i n_a\partial_i n_a+\partial_t n_a(\tau \partial_tn+ n)-D\partial_i n_a \partial_i n\;.
			\end{equation}
	Note that the above Lagrangian was constructed in the framework of Martin-Siggia-Rose (MSR) in Ref.~\cite{Abbasi:2022aao}. 
    However, later on, it was shown in Ref.~\cite{Jain:2023obu} that this indeed gives the KMS-invariant effective action. 
    In other words, the above Lagrangian is exactly what one finds in the Schwinger-Keldysh framework for the causal diffusion. 
    However, the KMS symmetry is not manifest, and the way the fields couple to external sources is not as simple as in the acausal theory of diffusion \cite{Chen-Lin:2018kfl}. 
    In the latter case, one just needs to impose $\partial_{\mu}n_a\rightarrow \partial_{\mu}n_a+A_{a,\mu}$, and $\partial_{\mu}n_r\rightarrow \partial_{\mu}n_r+A_{r,\mu}$. 
    For the causal theory, however, as shown in Ref.~\cite{Jain:2023obu} (see Appendix A therein), the Lagrangian in the presence of noise sources is given by
	\begin{equation}\label{L_2_new}
		\mathcal{L}^{\text{sourced}}_2=iT \sigma  \partial_i n_a\partial_i n_a+\partial_t n_a(\tau \partial_tn+ n)+A_{a,t}n-D\partial_i n_a \partial_i n+\sigma \partial_i n_a\partial_i A_{r,t}\;.
	\end{equation}
		The correlation function we are interested in is $G_{J^tJ^t}$ and is given by (we denote $p\equiv(\omega, \mathbf{k})$):
\begin{equation}\label{G_rr}
	G_{J^tJ^t}(p)=\frac{1}{i^2}\frac{\delta^2 \ln Z}{\delta A_{at}(-p)\delta A_{at}(p)}\;.
\end{equation}
where $Z[A]=\int \mathcal{D}n\mathcal{D}n_a e^{i S_{\text{eff}}[n,n_a; A]}$. 
Thus, we only need to turn on an $A_{a,t}$ source. 
There is only one term in the effective action that could couple to such an external source. 
We find
\begin{equation}\label{gauged_action}
	S_{\text{eff}}[A]= \int \mathrm{d}^{d+1}x\big(\mathcal{L}_{2}+ A_{a,t}n \big)\;.
	\end{equation}
	Using this, Eq.~\eqref{G_rr} gives Eq.~\eqref{correlation}, i.e., $G_{J^tJ^t}\equiv G_{nn}$.
	
The SK-EFT has recently been constructed to study a variety of systems. 
Notable examples are:
An SK-EFT associated with kinetic theory in the relaxation-time approximation was constructed in Ref.~\cite{Abbasi:2024pwz}. 
This work calculates the two- and three-point correlation functions of the density to all orders in the derivative expansion.
The computation of three-point functions from SK-EFT at leading order in the derivative expansion was performed earlier in Ref.~\cite{Delacretaz:2023ypv}.
SK-EFTs have also been constructed for light in medium \cite{Salcedo:2024nex}, and systems with approximate symmetries \cite{Delacretaz:2021qqu,Hongo:2024brb}.
Extensions include dissipative fluids with one-form symmetries \cite{Vardhan:2024qdi}, applications in the cosmological context \cite{Ota:2024mps}, study of off-equilibrium fluctuations \cite{Sogabe:2021svv}, and also hydrodynamics without Lorentz boosts \cite{Armas:2020mpr}.
	\section{Hydrodynamic loops from SK-EFT}
	\label{loop}
    Loop calculations for the theory of diffusion are addressed in several works in the literature. 
    In addition to Refs.~\cite{Kovtun:2012rj, Kovtun:2011np,Chen-Lin:2018kfl, Michailidis:2023mkd, Abbasi:2022aao} the effects of higher loops on the late-time behavior of thermal correlators are explored in Ref.~\cite{Delacretaz:2020nit} (see also Ref.~\cite{Grozdanov:2024fle}). 
    In Ref.~\cite{Jain:2020zhu}, the impact of non-Gaussian noise terms on the late-time behavior of correlators is investigated. 
    Reference~\cite{Abbasi:2021fcz} examines the role of loop effects on the energy diffusion through an SYK chain. 
    Loop calculations have also been utilized to establish bounds on thermalization timescales, as discussed in Ref.~\cite{Delacretaz:2023pxm}.

To calculate the effect of loops, one must first construct the interaction part of the Lagrangian. 
This is achieved by allowing the parameters in Eq.~\eqref{L_2_new} to depend on the density field $n$, or equivalently, the chemical potential $\mu$. 
As demonstrated in Ref.~\cite{Jain:2023obu}, preserving the KMS symmetry at nonlinear order requires the condition $\frac{\tau'(n)}{\tau}=\frac{\sigma'(n)}{\sigma}$.
The effective field theory considered in this work corresponds to a special case that satisfies this condition, in which both $\tau$ and $\sigma$ are taken to be constant.
Consequently, the only source of nonlinearity in this setup arises from the diffusivity $D\equiv D(n)$. 
This explains why the coupling constant $\lambda_{\sigma}$ in Eq.~(7-A) of Table.~\ref{Table} does not appear in Eq.~(7-B).

Interestingly, none of the physical quantities associated with Theory A discussed in this paper depend on $\lambda_{\sigma}$. 
This supports the validity of the comparisons made throughout this work, at least for systems with $\tau=$~const. and $\sigma=$~const.. A
comprehensive analysis of the more general case, where all parameters are allowed to vary, will be left to future work.

     It is easy to show that in any number of dimensions, the interactions in both Theory A and Theory B are irrelevant in the limit $k\rightarrow 0$ \cite{Machta:1984,Michailidis:2023mkd}. 
     This is because in this limit $\omega\sim k^2$, and thus $n\sim n_a\sim k^{d/2}$.\footnote{We take $\partial_i\sim k$, and $\partial_t\sim k^2$; therefore $\mathcal{L}\sim k^{d+2}$. Thus, from the first or third term in Eq.~\eqref{L_2_new}, we obtain $n\sim n_a \sim k^{d/2}$. Note that the second term is sub-leading at $k\rightarrow 0$.}
     Consequently, the interaction term is suppressed by a higher power of $k$ compared to the quadratic term.\footnote{This conclusion does not hold for momentum-conserving systems. In such systems, the interactions are irrelevant only for $d>2$. At $d=2$ they are marginal, while at $d=1$ they are relevant, and the theory flows to a
	new dissipative IR fixed point with dynamic exponent $z=3/2$ \cite{Delacretaz:2020jis}.} Therefore, the perturbative expansion is always controlled and one can systematically investigate the effects of loops on the late-time diffusive  transport. 
    This approach has been applied in Refs.~\cite{Chen-Lin:2018kfl} and \cite{Abbasi:2022aao} to calculate the loop corrections in Theory A and Theory B, respectively. 

In the limit $\tau D k^2\gg1$, which is specific to Theory B, $\omega\sim 1/\tau$ and the interactions become relevant. 
This can be simply seen by taking the limit $\tau D k^2\gg 1$ in $\delta D$ in Eq.~(9-B) in Table~\ref{Table}. 
The result is $\delta D=\frac{\lambda_D^2  T \chi}{16 D^{3/2}\tau^{1/2}}(- i \omega \tau)$,  which is non-vanishing for $\omega\sim 1/\tau$. 
The only way for the EFT to remain predictive in this limit is to fine-tune the coupling constant. 
As shown in Eq.~\eqref{coupling}, the coupling constants $D^{(m)}=\mathrm{d}^mD/\mathrm{d}n^m$ are inversely proportional to $\chi^m$. 
Thus, in the limit $\tau D k^2\gg1$, the perturbative expansion may remain controlled if the susceptibility is sufficiently large. 
This assumption has been made in Ref.~\cite{Abbasi:2022aao} to calculate the loop corrections in Theory B (see below).

	 The loop corrections to $G_{n n_a}$ can be written as follows ($p=(\omega, \mathbf{k}))$:
	\begin{equation}\label{Sigma}
		G_{n n_a}(p)=\,G^{(0)}_{n n_a}(p)+G^{(0)}_{n n_a}(p)(-\Sigma(p))G^{(0)}_{n n_a}(p)\,+\ldots
		=\,\frac{1}{\omega+ i D k^2- i \tau\, \omega^2+\,\Sigma(\omega, \textbf{k})} \; ,
	\end{equation}
	with $\Sigma$ being the self-energy appearing in the retarded Green's function~(7). 
	It turns out that there are at most two one-loop diagrams contributing in both Theories A and B (see Ref.~\cite{Michailidis:2023mkd} for Theory A and Ref.~\cite{Abbasi:2022aao} for Theory B):
	\begin{equation}\label{Sigma_0}
		\begin{split}
			G^{(0)}_{n n_a}(p)\Sigma(p)G^{(0)}_{n n_a}(p)=&\,
			\scalebox{0.53}{	\begin{tikzpicture}[baseline=(a.base)]
					\begin{feynman}
						\vertex (a) ;
						\vertex [right=of a] (a1) ;
						\vertex [right=of a1] (a2);
						\vertex [above right=of a2] (a3);
						\vertex [below right=of a2] (a4);
						\vertex [above right=of a4] (a5); 
						\vertex [ right=of a5] (a6);
						\vertex [ right=of a6] (a7);
						\diagram* {
							(a) --[very thick](a1)-- [boson,very thick] (a2) -- [quarter left, very thick] (a3)--[boson,quarter left,very thick](a5)--[quarter left, very thick](a4)--[quarter left,very  thick](a2) ,
							(a5) -- [very thick] (a6)--  [boson,very thick](a7),
						};
					\end{feynman}
			\end{tikzpicture}}\, 	
			+
		\scalebox{0.53}{	\begin{tikzpicture}[baseline=(a.base)]
		\begin{feynman}
			\vertex (a) ;
			\vertex [right=of a] (a1) ;
			\vertex [right=of a1] (a2);
			\vertex [above right=of a2] (a3);
			\vertex [below right=of a2] (a4);
			\vertex [above right=of a4] (a5); 
			\vertex [ right=of a5] (a6);
			\vertex [ right=of a6] (a7);
			\diagram* {
				(a) --[very thick](a1)-- [boson,very thick] (a2) -- [quarter left, very thick] (a3)--[quarter left,boson,very thick](a5)--[quarter left,boson,very thick](a4)--[quarter left, very thick](a2) ,
				(a5) --  [very thick](a6)--  [boson,very thick](a7),
			};\;.
		\end{feynman}
	\end{tikzpicture}} \, ,
		\end{split}
	\end{equation}
	where $G^{(0)}_{n n}(p)$ (see Eq.~\eqref{correlation}) and  $G^{(0)}_{n n_a}(p)$ correspond to solid and half-solid half-wavy lines, respectively. 
    In Theory B, 
	\begin{equation}\label{G0_nn_a}
	G^{(0)}_{nn_a}=\frac{1}{\omega+ i D k^2- i \tau \omega^2}\;.
	\end{equation}
Setting $\tau=0$, we get the corresponding expression in Theory A.
The self-energy is given by the following integral
\begin{equation}\label{sigma}
		\Sigma(p)=\,\lambda_D^2\,k^2\int_{p'}k'^2G^{(0)}_{n_a n}(p')G^{(0)}_{n n}(p'+p)+i \chi T\lambda_D \lambda_{\sigma} k^2\int_{p'}\,\,(k'^2+ \mathbf{k}\cdot \mathbf{k}')G^{(0)}_{n_a n}(p')G^{(0)}_{n n_a}(p+p')\;.
\end{equation}
Note that in Theory B, $\lambda_{\sigma}=0$ \cite{Jain:2023obu}.
To calculate $\delta \tau$ and $\delta D$ in Eqs.~(8-A) and (8-B), one way is to first calculate the corrected $G_{nn}$ and then use the fluctuation-dissipation theorem (FDT) to find $G_{R}$. 
Let us first parameterize $G_{R}$ as follows:
\begin{equation}\label{G_R}
	G_{R}(\omega,k)=\,\frac{i\,\big[\sigma+\delta \sigma(p)\big]\,k^2}{- i \tau \omega^2+ \omega +i D k^2+ \Sigma(p) }\;.
\end{equation}
Then the FDT implies that	
	\begin{equation}\label{G_{nn}}
		G_{nn}(\omega,\mathbf{k})=\,\frac{	N(p)}{\omega^2+D^2 k^4+2\omega\,\text{Re}\,\Sigma(p)+2(D k^2- \tau \omega^2)\,\text{Im}\,\Sigma(p)}\; ,
	\end{equation}
	with the numerator $N(p)$ given by
	\begin{equation}\label{C}
		N(p)=2 T \chi D  k^2\left[1+\frac{\text{Re}\delta \sigma(p)}{\sigma}+\frac{D k^2-\tau \omega^2}{\omega}\frac{\text
			{Im}\delta \sigma(p)}{\sigma}+\frac{\text{Re}\Sigma(p)}{\omega}\right] \; .
	\end{equation}
	Now the idea of Ref.~\cite{Chen-Lin:2018kfl} (developed in Ref.~\cite{Abbasi:2022aao} for the case of Theory B) is to calculate the numerator separately. 
    It turns out that
	\begin{align}
			G^{(0)}_{n n_a}(p)(-N(p))G^{(0)}_{n_a n}(p)=
			\feynmandiagram[scale=0.5,transform shape] [inline=(a.base),horizontal=a to b] {a --[]c-- []b};
		+
\scalebox{0.5}{	\begin{tikzpicture}[baseline=(a.base)]
		\begin{feynman}
			\vertex (a) ;
			\vertex [right=of a] (a1) ;
			\vertex [right=of a1] (a2);
			\vertex [above right=of a2] (a3);
			\vertex [below right=of a2] (a4);
			\vertex [above right=of a4] (a5); 
			\vertex [ right=of a5] (a6);
			\vertex [ right=of a6] (a7);
			\diagram* {
				(a) -- [ very thick](a1)-- [boson,very thick] (a2) -- [quarter left, very thick] (a3)--[quarter left, very thick](a5)--[quarter left, very thick](a4)--[quarter left, very thick](a2) ,
				(a5) -- [boson,very thick] (a6)--   [very thick](a7),
			};
		\end{feynman}
\end{tikzpicture}}
			+
			\left [\, \scalebox{0.5}{	\begin{tikzpicture}[baseline=(a.base)]
					\begin{feynman}
						\vertex (a) ;
						\vertex [right=of a] (a1) ;
						\vertex [right=of a1] (a2);
						\vertex [above right=of a2] (a3);
						\vertex [below right=of a2] (a4);
						\vertex [above right=of a4] (a5); 
						\vertex [ right=of a5] (a6);
						\vertex [ right=of a6] (a7);
						\diagram* {
							(a) --[very thick](a1)-- [boson,very thick] (a2) -- [quarter left, very thick] (a3)--[quarter left,boson,very thick](a5)--[quarter left, very thick](a4)--[quarter left, very thick](a2) ,
							(a5) -- [boson,very thick] (a6)--  [very thick](a7),
						};
					\end{feynman}
			\end{tikzpicture}}
			+\text{c.c.} \right]\; ,
	\end{align}
	with
	\begin{equation}\label{C_int}
			N(p)=2 T \chi D  k^2 +
			\frac{1}{2}\lambda_D^2k^4\int_{p'}G^{(0)}_{nn}(p')G^{(0)}_{nn}(p-p')
			+i \chi T \lambda_{\sigma}\lambda_D k^2
			\int_{p'}\mathbf{k}\cdot \mathbf{k}'G^{(0)}_{n_a n}(p')G^{(0)}_{nn}(p+p')\; .
	\end{equation}
	Note that in Theory B, $\lambda_{\sigma}=0$ \cite{Jain:2023obu}. 
    Calculating the expressions \eqref{sigma} and \eqref{C_int}, then we can read off $\delta \sigma(p)$ from Eq.~\eqref{C}. 
    We refer the reader to Ref.~\cite{Abbasi:2022aao} for the details of performing the integrals.  
    The last step is to rewrite the response function \eqref{G_R} in the following form:
	\begin{equation}\label{G_R_c}
		G^{R}_{nn}(\omega,\mathbf{k})=\,\frac{i\,\sigma\,k^2}{- i \left[\tau+\delta \tau(\omega,\textbf{k})\right] \omega^2+ \omega +i \left[D+\delta D(\omega,\textbf{k})\right] k^2}\;.
	\end{equation}
	The one-loop corrections to $\tau$ and $D$ appearing in this expression are given by Eqs.~(9-A) and (9-B) in Table~\ref{Table}.
	
	
	\section{Singularities from on-shell conditions}
    \label{singularities}
Let us recall that in order to determine the branch points of the one-loop correlation functions, it suffices to use the tree-level dispersion relations, i.e., the linear-response modes given by Eqs.~(3-A) and (3-B). In what follows, we will identify the branch points in the lower half-plane of the correlation function $G_{nn_{B}}$.
 
Let us define 
 \begin{equation}\label{omega_ij}
 	\omega_{ij}(\textbf{k},\textbf{k}')=\,\omega_{i}(\textbf{k}')+\omega_{j}(\textbf{k}-\textbf{k}')\;.
 \end{equation}
For $i=j=1$ or $i=j=2$, we obtain 
 \begin{equation}\label{omega_11}
 	\omega_{11/22}(\textbf{k},\textbf{k}')=\,-\frac{i}{2\tau}\left[1\mp\sqrt{1- 4 \tau D \textbf{k}'^2}+1\mp\sqrt{1- 4 \tau D (\textbf{k}-\textbf{k}')^2}\right]\,.
 \end{equation}
    Solving $\partial_{\textbf{k}'}\omega_{11/22}(\textbf{k},\textbf{k}')=0$, we obtain $\textbf{k}'=\frac{1}{2}\textbf{k}$. Substituting this into Eq.~\eqref{omega_11}, we find
 \begin{equation}\label{omega_11a}
 	\tilde{\omega}_{11/22}=\omega_{11/22}\left(\textbf{k},\textbf{k}'=\frac{1}{2}\textbf{k}\right)=\,-\frac{i}{\tau}\big(1\mp\sqrt{1-\tau D k^2}\big)\,.
 \end{equation}
For $i=1,\,j=2$ or vice versa, we have 
 \begin{equation}\label{omega_12}
 	\omega_{12/21}(\textbf{k},\textbf{k}')=\,-\frac{i}{2\tau}\left[1\mp\sqrt{1- 4 \tau D \textbf{k}'^2}+1\pm\sqrt{1- 4 \tau D (\textbf{k}-\textbf{k}')^2}\right]\,.
 \end{equation}
    Solving $\partial_{\textbf{k}'}\omega_{12/21}(\textbf{k},\textbf{k}')=0$, we obtain $k'=+\infty$. 
    Expanding Eq.~\eqref{omega_12} around $k'\rightarrow+\infty$, we obtain
 \begin{equation}\label{omega_12a}
\tilde{\omega}_{12/21}=-\frac{i}{\tau}\pm \frac{\sqrt{\tau D k^2 }}{\tau}\,.
 \end{equation}
See also Sec.~4.2.2 in Ref.~\cite{Abbasi:2022rum} for an alternative derivation based on Landau-loop analysis.

	\section{Contour integrals}
	\subsection{The overdamped limit}
    \label{App_overdamped}
    To perform the Fourier integral  $\int \frac{\mathrm{d}\omega}{2\pi}G_{nn}(\omega, \mathbf{k})\,e^{-i \omega t}$ for $t>0$ for Eq.~\eqref{G_nn_A_corr}, we consider the analytic structure of $G_{nn}(\omega, \mathbf{k})$. In the limit $\tau \omega\sim \tau D k^2 \ll 1$, it is illustrated in left panel of Fig.~\ref{contour_long_time_tail} in the lower half of the complex $\omega$-plane.

  Now taking $\omega=\tilde{\omega}+i D k^2 z$, where $\tilde{\omega}=-\frac{i}{2}D k^2$, the Fourier integral takes the following form 
\begin{equation}   \label{delta_Gnn_A} 
	\delta G_{{nn}_A}(t, \mathbf{k})=\frac{\lambda_{D}^2}{4 \sqrt{2} D^2}T^2 \chi^2\,|k|\,e^{- \frac{1}{2}D k^2 t}\int_{-\infty}^{0} \frac{i \mathrm{d} z}{2\pi}\,\text{Disc}\frac{e^{Dk^2 z t}}{\sqrt{z}\big(z +\frac{1}{2}\big)^2}\;,
\end{equation}
where $\text{Disc} f(z)=\lim\limits_{\epsilon\to 0}\left[f(z+ i \epsilon)-f(z-i \epsilon)\right]$.
Evaluating the integral,  we find the result given in Eq.~\eqref{scale_A}.  
This was first found for Theory A in Ref.~\cite{Michailidis:2023mkd}.

\subsection{The underdamped limit}
\label{App_underdamped}
Let us rewrite Eq.~\eqref{G_nn_B_corr} in the following form:
\begin{equation}\label{delta_Gnn_B}
	\delta G_{{nn}_{B}}(\omega,\mathbf{k})=
\frac{\lambda_D^2 T^2 \chi^2}{16D^{1/2}}  k^4 
\left[ \frac{2 + \tau D k^2 - \tau \omega(3i+\tau \omega)}{\tau D k^2 + (1 -i \tau \omega)^2} \right]^2 \frac{\sqrt{\frac{\tau D k^2 + (1 -i \tau \omega)^2}{D k^2 - i\omega (2 - i\tau  \omega)}}}{(Dk^2 - i \omega - \tau \omega^2)^2}\;. 
	\end{equation}
We can replace each of the square polynomials of $\omega$ with $(\omega-\omega_1)(\omega-\omega_2)$, where $\omega_{1,2}$ are the two roots of the expression. 
In the limit $\tau D k^2\gg1$, the roots of the numerator and denominator in the square-root expression converge, resulting in the disappearance of the square-root singularity. 
This corresponds to the contraction of the two branch cuts, as illustrated in the bottom-right panel of Fig.~\ref{analytic}. 
We find
	\begin{align}\label{delta_Gnn_Ba}
		\delta G_{{nn}_B}(\omega,\mathbf{k})& =\frac{\lambda_D^2 T^2 \chi^2}{16 \sqrt{D \tau^3}} k^4 \frac{1}{\big(\omega+\frac{i}{\tau}+\frac{\sqrt{\tau D k^2}}{\tau}\big)^2\big(\omega+\frac{i}{\tau}-\frac{\sqrt{\tau D k^2}}{\tau}\big)^2}  \nonumber \\
        & \times \frac{\big(\omega+\frac{3i}{2\tau}+\frac{\sqrt{\tau D k^2}}{\tau}\big)^2\big(\omega+\frac{3i}{2\tau}-\frac{\sqrt{\tau D k^2}}{\tau}\big)^2}{\big(\omega+\frac{i}{2\tau}+\frac{\sqrt{\tau D k^2}}{\tau}\big)^2\big(\omega+\frac{i}{2\tau}-\frac{\sqrt{\tau D k^2}}{\tau}\big)^2}\;.
	\end{align}
	Now $\delta G_{nn}(\omega,\mathbf{k})$ has four poles, each of which is a second-order pole. 
    Using the residue theorem for the analytic function $h$:\footnote{The contour is counter-clockwise in the complex $z$-plane.}
\begin{equation}\label{residue}
\oint \mathrm{d}z \frac{h(z)}{(z-z_0)^2}=2\pi i \lim_{z\rightarrow z_0}\frac{\mathrm{d}}{\mathrm{d}z}h(z)\;,
\end{equation}
one can calculate $\int_{-\infty}^{+\infty}\frac{\mathrm{d}\omega}{2\pi}\delta G_{nn}(\omega,\mathbf{k})e^{- i \omega t}$ by closing the contour in the lower half of the complex $\omega$-plane. 
To leading order in $\tau D k^2$, the result is given by Eq.~\eqref{real_B}.
	\bibliographystyle{utphys}

\begin{thebibliography}{100}

\bibitem{Fick:1855}
A.~Fick,
``\"Uber Diffusion,''
Ann.~Phys.\textbf{94} (1855) no.1, 59-86.


\bibitem{Cattaneo:1958}
C.~R.~Cattaneo,
``Sur une forme de l'équation de la chaleur éliminant le paradoxe d'une propagation instantanée,''
Comptes Rendus \textbf{247} (1958) no.4, 431.


\bibitem{Koide:2006ef}
T.~Koide, G.~S.~Denicol, P.~Mota and T.~Kodama,
``Relativistic dissipative hydrodynamics: A Minimal causal theory,''
Phys. Rev. C \textbf{75} (2007), 034909
[arXiv:hep-ph/0609117 [hep-ph]].


\bibitem{Bemfica:2020zjp}
F.~S.~Bemfica, M.~M.~Disconzi and J.~Noronha,
``First-Order General-Relativistic Viscous Fluid Dynamics,''
Phys. Rev. X \textbf{12} (2022) no.2, 021044
[arXiv:2009.11388 [gr-qc]].

\bibitem{Jain:2023obu}
A.~Jain and P.~Kovtun,
``Schwinger-Keldysh effective field theory for stable and causal relativistic hydrodynamics,''
JHEP \textbf{01} (2024), 162
[arXiv:2309.00511 [hep-th]].

\bibitem{Pu:2009fj}
S.~Pu, T.~Koide and D.~H.~Rischke,
``Does stability of relativistic dissipative fluid dynamics imply causality?,''
Phys. Rev. D \textbf{81} (2010), 114039
[arXiv:0907.3906 [hep-ph]].


\bibitem{Gavassino:2024ufs}
L.~Gavassino, N.~Abboud, E.~Speranza and J.~Noronha,
``First-order relativistic hydrodynamics with an information current,''
Phys. Rev. D \textbf{109} (2024) no.8, 085013
[arXiv:2401.13852 [nucl-th]].



\bibitem{Grozdanov:2018fic}
S.~Grozdanov, A.~Lucas and N.~Poovuttikul,
``Holography and hydrodynamics with weakly broken symmetries,''
Phys. Rev. D \textbf{99} (2019) no.8, 086012
[arXiv:1810.10016 [hep-th]].



\bibitem{Stephanov:2017ghc}
M.~Stephanov and Y.~Yin,
``Hydrodynamics with parametric slowing down and fluctuations near the critical point,''
Phys. Rev. D \textbf{98} (2018) no.3, 036006
[arXiv:1712.10305 [nucl-th]].


	\bibitem{Brattan:2024dfv}
	D.~K.~Brattan, M.~Matsumoto, M.~Baggioli and A.~Amoretti,
	``Relaxed hydrodynamic theory of electrically driven non-equilibrium steady states,''
	Phys. Rev. Res. \textbf{6}, 043097
	[arXiv:2404.05568 [cond-mat.stat-mech]].


\bibitem{Denicol:2011fa}
G.~S.~Denicol, J.~Noronha, H.~Niemi and D.~H.~Rischke,
``Origin of the Relaxation Time in Dissipative Fluid Dynamics,''
Phys. Rev. D \textbf{83} (2011), 074019
[arXiv:1102.4780 [hep-th]].


\bibitem{Denicol:2012cn}
G.~S.~Denicol, H.~Niemi, E.~Molnar and D.~H.~Rischke,
Phys. Rev. D \textbf{85}, 114047 (2012)
[erratum: Phys. Rev. D \textbf{91}, no.3, 039902 (2015)]
[arXiv:1202.4551 [nucl-th]].

\bibitem{Brown:2019}
P.~T.~Brown, D.~Mitchell, M.~Guardado-Sanchez, R.~Kondov, E.~N.~Demler, W.~S.~Bakr,
``Science 363, 379 (2019),''
arXiv:1802.09456 [cond-mat.quant-gas].



\bibitem{Du:2021zqz}
L.~Du, X.~An and U.~Heinz,
``Baryon transport and the QCD critical point,''
Phys. Rev. C \textbf{104} (2021) no.6, 064904
[arXiv:2107.02302 [hep-ph]].


\bibitem{Chen-Lin:2018kfl}
X.~Chen-Lin, L.~V.~Delacr\'etaz and S.~A.~Hartnoll,
``Theory of diffusive fluctuations,''
Phys. Rev. Lett. \textbf{122} (2019) no.9, 091602
[arXiv:1811.12540 [hep-th]].

\bibitem{Abbasi:2022aao}
N.~Abbasi, M.~Kaminski and O.~Tavakol,
``Theory of Nonlinear Diffusion with a Physical Gapped Mode,''
Phys. Rev. Lett. \textbf{132} (2024) no.13, 131602
[arXiv:2212.11499 [hep-th]].


\bibitem{Heller:2020uuy}
M.~P.~Heller, A.~Serantes, M.~Spali\'nski, V.~Svensson and B.~Withers,
``Hydrodynamic gradient expansion in linear response theory,''
Phys. Rev. D \textbf{104} (2021) no.6, 066002
[arXiv:2007.05524 [hep-th]].



\bibitem{Dash:2022xkz}
A.~Dash, M.~Shokri, L.~Rezzolla and D.~H.~Rischke,
``Charge diffusion in relativistic resistive second-order dissipative magnetohydrodynamics,''
Phys. Rev. D \textbf{107}, no.5, 056003 (2023)
[arXiv:2211.09459 [nucl-th]].

\bibitem{Lier:2025wfw}
R.~Lier, A.~Jain, J.~Armas and O.~Porth,
``Resistive relativistic magnetohydrodynamics without Amperes Law,''
[arXiv:2501.04638 [astro-ph.HE]].

\bibitem{Romatschke:2017ejr}
P.~Romatschke and U.~Romatschke,
``Relativistic Fluid Dynamics In and Out of Equilibrium,''
Cambridge University Press, 2019,
ISBN 978-1-108-48368-1, 978-1-108-75002-8
[arXiv:1712.05815 [nucl-th]].

\bibitem{Romatschke:2009im}
P.~Romatschke,
``New Developments in Relativistic Viscous Hydrodynamics,''
Int. J. Mod. Phys. E \textbf{19} (2010), 1-53
[arXiv:0902.3663 [hep-ph]].


	
\bibitem{Liu:2018kfw}
H.~Liu and P.~Glorioso,
``Lectures on non-equilibrium effective field theories and fluctuating hydrodynamics,''
PoS \textbf{TASI2017}, 008 (2018)
[arXiv:1805.09331 [hep-th]].



\bibitem{Glorioso:2017fpd}
P.~Glorioso, M.~Crossley and H.~Liu,
``Effective field theory of dissipative fluids (II): classical limit, dynamical KMS symmetry and entropy current,''
JHEP \textbf{09} (2017), 096
[arXiv:1701.07817 [hep-th]].


\bibitem{Jensen:2018hhx}
K.~Jensen, R.~Marjieh, N.~Pinzani-Fokeeva and A.~Yarom,
``An entropy current in superspace,''
JHEP \textbf{01} (2019), 061
[arXiv:1803.07070 [hep-th]].

\bibitem{Jensen:2018hse}
K.~Jensen, R.~Marjieh, N.~Pinzani-Fokeeva and A.~Yarom,
``A panoply of Schwinger-Keldysh transport,''
SciPost Phys. \textbf{5} (2018) no.5, 053
[arXiv:1804.04654 [hep-th]].


\bibitem{Haehl:2018lcu}
F.~M.~Haehl, R.~Loganayagam and M.~Rangamani,
``Effective Action for Relativistic Hydrodynamics: Fluctuations, Dissipation, and Entropy Inflow,''
JHEP \textbf{10} (2018), 194
[arXiv:1803.11155 [hep-th]].

\bibitem{Grozdanov:2013dba}
S.~Grozdanov and J.~Polonyi,
``Viscosity and dissipative hydrodynamics from effective field theory,''
Phys. Rev. D \textbf{91} (2015) no.10, 105031
[arXiv:1305.3670 [hep-th]].

\bibitem{Mullins:2025vqa}
N.~Mullins, M.~Hippert and J.~Noronha,
``Effective Action for Relativistic Hydrodynamics from the Crooks Fluctuation Theorem,''
Phys. Rev. Lett. \textbf{134} (2025) no.23, 232302
[arXiv:2501.04637 [nucl-th]].


\bibitem{Basar:2024srd}
G.~Basar,
``Recent developments in relativistic hydrodynamic fluctuations,''
[arXiv:2410.02866 [hep-th]].

				\bibitem{Landau_1}
L. D. ~Landau and E. M. ~Lifshitz,
``Statistical Physics Part 1,''
Course of Theoretical Physics,
Vol 5 (Elsevier Science, 2013).


\bibitem{Michailidis:2023mkd}
A.~A.~Michailidis, D.~A.~Abanin and L.~V.~Delacr\'etaz,
``Corrections to Diffusion in Interacting Quantum Systems,''
Phys. Rev. X \textbf{14} (2024) no.3, 031020
[arXiv:2310.10564 [cond-mat.stat-mech]].


\bibitem{Delacretaz:2020jis}
L.~V.~Delacretaz and P.~Glorioso,
``Breakdown of Diffusion on Chiral Edges,''
Phys. Rev. Lett. \textbf{124} (2020) no.23, 236802
[arXiv:2002.08365 [cond-mat.str-el]].





\bibitem{Mullins:2023ott}
N.~Mullins, M.~Hippert, L.~Gavassino and J.~Noronha,
``Relativistic hydrodynamic fluctuations from an effective action: Causality, stability, and the information current,''
Phys. Rev. D \textbf{108} (2023) no.11, 116019
[arXiv:2309.00512 [hep-th]].


\bibitem{Glorioso:2018mmw}
P.~Glorioso, M.~Crossley and H.~Liu,
[arXiv:1812.08785 [hep-th]].



\bibitem{Bu:2021clf}
Y.~Bu, M.~Fujita and S.~Lin,
``Ginzburg-Landau effective action for a fluctuating holographic superconductor,''
JHEP \textbf{09} (2021), 168
[arXiv:2106.00556 [hep-th]].



\bibitem{Liu:2024tqe}
Y.~Liu, Y.~W.~Sun and X.~M.~Wu,
``Holographic Schwinger-Keldysh effective field theories including a non-hydrodynamic mode,''
[arXiv:2411.16306 [hep-th]].


\bibitem{Ahn:2025odk}
Y.~Ahn, M.~Baggioli, Y.~Bu, M.~Matsumoto and X.~Sun,
``Simple holograpic dual of the Maxwell-Cattaneo model,''
[arXiv:2506.00926 [hep-th]].

\bibitem{Kapusta:2014dja}
J.~I.~Kapusta and C.~Young,
``Causal Baryon Diffusion and Colored Noise,''
Phys. Rev. C \textbf{90} (2014) no.4, 044902
[arXiv:1404.4894 [nucl-th]].

\bibitem{Landau:1959}
L.~D.~Landau,
``On analytic properties of vertex parts in quantum field theory,''
Nuclear Physics \textbf{13} (1959), 181-192.

\bibitem{Eden:1966}
R.~J.~Eden, P.~V.~Landshoff, D.~I.~Olive, and J.~C.~Polkinghorne,
``The Analytic S-Matrix,''
Cambridge University Press, 1966.


\bibitem{Bourjaily:2020wvq}
J.~L.~Bourjaily, H.~Hannesdottir, A.~J.~McLeod, M.~D.~Schwartz and C.~Vergu,
``Sequential Discontinuities of Feynman Integrals and the Monodromy Group,''
JHEP \textbf{01} (2021), 205
[arXiv:2007.13747 [hep-th]].


\bibitem{Huber:2023uzd}
M.~Q.~Huber, W.~J.~Kern and R.~Alkofer,
``How to Determine the Branch Points of Correlation Functions in Euclidean Space II: Three-Point Functions,''
Symmetry \textbf{15} (2023) no.2, 414
[arXiv:2302.01350 [hep-ph]].



\bibitem{Abbasi:2022rum}
N.~Abbasi, A.~Davody and S.~Tahery,
``Correlation functions in stable first-order relativistic hydrodynamics,''
Phys. Rev. D \textbf{109} (2024) no.3, 036006
[arXiv:2212.14619 [hep-th]].




\bibitem{Glorioso:2022poi}
P.~Glorioso and S.~A.~Hartnoll,
``Joule heating in bad and slow metals,''
SciPost Phys. \textbf{13} (2022) no.4, 095
[arXiv:2202.00689 [cond-mat.str-el]].

\bibitem{Chagnet:2023xsl}
N.~Chagnet and K.~Schalm,
``Hydrodynamics of a relativistic charged fluid in the presence of a periodically modulated chemical potential,''
SciPost Phys. \textbf{16} (2024) no.1, 028
[arXiv:2303.17685 [cond-mat.str-el]].

\bibitem{Varney:2009}
C.~N.~Varney, C.-R.~Lee, Z.~J.~Bai, S.~Chiesa, M.~Jarrell and R.~T.~Scalettar,
``Quantum Monte Carlo study of the two-dimensional fermion Hubbard Model,''
Phys.\ Rev.\ B \textbf{80} (2009), 075116
[arXiv:0906.4311 [cond-mat.str-el]].



%


\bibitem{Son:2007vk}
D.~T.~Son and A.~O.~Starinets,
``Viscosity, Black Holes, and Quantum Field Theory,''
Ann. Rev. Nucl. Part. Sci. \textbf{57} (2007), 95-118
[arXiv:0704.0240 [hep-th]].



\bibitem{Oei:2024zyx}
N.~Oei, N.~Krenz, H.~van Hees, C.~Greiner and J.~M.~Torres-Rincon,
``Formation, dissociation, and regeneration of charmonia within microscopic Langevin simulations,''
Phys. Rev. D \textbf{111} (2025) no.7, 074012
[arXiv:2410.19619 [hep-ph]].


\bibitem{Son:2009vu}
D.~T.~Son and D.~Teaney,
``Thermal Noise and Stochastic Strings in AdS/CFT,''
JHEP \textbf{07} (2009), 021
[arXiv:0901.2338 [hep-th]].

\bibitem{Casalderrey-Solana:2007ahi}
J.~Casalderrey-Solana and D.~Teaney,
``Transverse Momentum Broadening of a Fast Quark in a N=4 Yang Mills Plasma,''
JHEP \textbf{04} (2007), 039
[arXiv:hep-th/0701123 [hep-th]].

\bibitem{Casalderrey-Solana:2006fio}
J.~Casalderrey-Solana and D.~Teaney,
``Heavy quark diffusion in strongly coupled N=4 Yang-Mills,''
Phys. Rev. D \textbf{74} (2006), 085012
[arXiv:hep-ph/0605199 [hep-ph]].



\bibitem{Rajagopal:2025ukd}
K.~Rajagopal, B.~Scheihing-Hitschfeld and U.~A.~Wiedemann,
``Dynamics of Heavy Quarks in Strongly Coupled $\mathcal{N}=4$ SYM Plasma,''
[arXiv:2501.06289 [hep-ph]].

\bibitem{Rajagopal:2025rxr}
K.~Rajagopal, B.~Scheihing-Hitschfeld and U.~A.~Wiedemann,
``A Universal Equilibration Condition for Heavy Quarks,''
[arXiv:2504.21139 [hep-ph]].


\bibitem{Lin:2023bli}
S.~Lin, Y.~Bu and C.~Lei,
``Non-Gaussianity from Schwinger-Keldysh effective field theory,''
Phys. Rev. D \textbf{109} (2024) no.3, 036018
[arXiv:2301.06703 [hep-th]].

\bibitem{Hohenberg:1977ym}
P.~C.~Hohenberg and B.~I.~Halperin,
``Theory of dynamic critical phenomena,''
Rev. Mod. Phys. \textbf{49} (1977), 435-479


\bibitem{Roth:2023wbp}
J.~V.~Roth and L.~von Smekal,
``Critical dynamics in a real-time formulation of the functional renormalization group,''
JHEP \textbf{10} (2023), 065
[arXiv:2303.11817 [hep-ph]].


\bibitem{Wu:2019qfz}
S.~Wu and H.~Song,
``Universal scaling of conserved charge in stochastic diffusion dynamics,''
Chin. Phys. C \textbf{43} (2019) no.8, 084103
[arXiv:1903.06075 [nucl-th]].


\bibitem{Sakai:2025xmo}
A.~Sakai, K.~Murase, H.~Fujii and T.~Hirano,
``Dynamical evolution of critical fluctuations with second-order baryon diffusion coupled to chiral condensate,''
[arXiv:2502.17791 [nucl-th]].


\bibitem{Abbasi:2024pwz}
N.~Abbasi and D.~H.~Rischke,
``Three-point functions from a Schwinger-Keldysh effective action, resummed in derivatives,''
[arXiv:2410.07929 [hep-th]].


\bibitem{Delacretaz:2023ypv}
L.~V.~Delacretaz and R.~Mishra,
``Nonlinear response in diffusive systems,''
SciPost Phys. \textbf{16} (2024) no.2, 047
[arXiv:2304.03236 [cond-mat.str-el]].


\bibitem{Salcedo:2024nex}
S.~A.~Salcedo, T.~Colas and E.~Pajer,
``An Open Effective Field Theory for light in a medium,''
[arXiv:2412.12299 [hep-th]].

\bibitem{Delacretaz:2021qqu}
L.~V.~Delacr\'etaz, B.~Gout\'eraux and V.~Ziogas,
``Damping of Pseudo-Goldstone Fields,''
Phys. Rev. Lett. \textbf{128} (2022) no.14, 141601
[arXiv:2111.13459 [hep-th]].

\bibitem{Hongo:2024brb}
M.~Hongo, N.~Sogabe, M.~A.~Stephanov and H.~U.~Yee,
``Schwinger-Keldysh effective action for hydrodynamics with approximate symmetries,''
[arXiv:2411.08016 [hep-th]].


\bibitem{Vardhan:2024qdi}
S.~Vardhan, S.~Grozdanov, S.~Leutheusser and H.~Liu,
``Effective field theories of dissipative fluids with one-form symmetries,''
[arXiv:2408.12868 [hep-th]].

\bibitem{Ota:2024mps}
A.~Ota,
``Fluctuation-dissipation relation in cosmic microwave background,''
JCAP \textbf{05} (2024), 062
[arXiv:2402.07623 [hep-th]].



\bibitem{Sogabe:2021svv}
N.~Sogabe and Y.~Yin,
``Off-equilibrium non-Gaussian fluctuations near the QCD critical point: an effective field theory perspective,''
JHEP \textbf{03} (2022), 124
[arXiv:2111.14667 [nucl-th]].


\bibitem{Armas:2020mpr}
J.~Armas and A.~Jain,
``Effective field theory for hydrodynamics without boosts,''
SciPost Phys. \textbf{11} (2021) no.3, 054
[arXiv:2010.15782 [hep-th]].



\bibitem{Kovtun:2012rj}
P.~Kovtun,
``Lectures on hydrodynamic fluctuations in relativistic theories,''
J. Phys. A \textbf{45} (2012), 473001
[arXiv:1205.5040 [hep-th]].


\bibitem{Kovtun:2011np}
P.~Kovtun, G.~D.~Moore and P.~Romatschke,
``The stickiness of sound: An absolute lower limit on viscosity and the breakdown of second order relativistic hydrodynamics,''
Phys. Rev. D \textbf{84} (2011), 025006
[arXiv:1104.1586 [hep-ph]].


\bibitem{Delacretaz:2020nit}
L.~V.~Delacretaz,
``Heavy Operators and Hydrodynamic Tails,''
SciPost Phys. \textbf{9} (2020) no.3, 034
[arXiv:2006.01139 [hep-th]].


\bibitem{Grozdanov:2024fle}
S.~Grozdanov, T.~Lemut, J.~Pelai\v{c} and A.~Soloviev,
``Analytic structure of diffusive correlation functions,''
Phys. Rev. D \textbf{110} (2024) no.5, 056053
[arXiv:2407.13550 [hep-th]].




\bibitem{Jain:2020zhu}
A.~Jain and P.~Kovtun,
``Late Time Correlations in Hydrodynamics: Beyond Constitutive Relations,''
Phys. Rev. Lett. \textbf{128} (2022) no.7, 7
[arXiv:2009.01356 [hep-th]].


\bibitem{Abbasi:2021fcz}
N.~Abbasi,
``Long-time tails in the SYK chain from the effective field theory with a large number of derivatives,''
JHEP \textbf{04} (2022), 181
[arXiv:2112.12751 [hep-th]].


\bibitem{Delacretaz:2023pxm}
L.~V.~Delacretaz,
``A Bound on Thermalization from Diffusive Fluctuations,''
[arXiv:2310.16948 [cond-mat.str-el]].

\bibitem{Machta:1984}
J.~Machta, M.~H.~Ernst, H.~van Beijeren, and J.~R.~Dorfman,
``Long Time Tails in Stationary Random Media II: Applications,''
Journal of Statistical Physics, \textbf{35} Nos. 3/4 (1984), 413-441.


\end{thebibliography}
	\providecommand{\href}[2]{#2}\begingroup\raggedright

\newpage{\pagestyle{empty}\cleardoublepage}

\end{document}